\documentclass[superscriptaddress,aps,pra,reprint,longbibliography]{revtex4-1}

\makeatletter
\def\@fnsymbol#1{\ensuremath{\ifcase#1\or \dagger\or *\or \ddagger\or
   \mathsection\or \mathparagraph\or \|\or **\or \dagger\dagger
   \or \ddagger\ddagger \else\@ctrerr\fi}}
\makeatother
    
\usepackage{amssymb,amsfonts,amsmath}
\usepackage{graphicx} 
\usepackage{bm}        
\usepackage{color}
\usepackage{float}
\usepackage{subfigure}
\usepackage{times}

\newcommand{\ket}[1]{$\left| #1 \right\rangle$}
\newcommand{\NN}[0]{N$_2^+$}
\newcommand{\Ca}[0]{Ca$^+$}
\newcommand{\Cadown}[0]{\ket{\downarrow}$_\text{Ca}$}
\newcommand{\Caup}[0]{\ket{\uparrow}$_\text{Ca}$}
\newcommand{\NNdown}[0]{\ket{\downarrow}$_{\text{N}_2}$}
\newcommand{\NNup}[0]{\{\ket{\uparrow}$_{\text{N}_2}$\}}
\newcommand{\am}[0]{\ket{\alpha}}
\newcommand{\bmm}[0]{\ket{\beta}}

\begin{document}

\title{Quantum non-demolition state detection and spectroscopy of single trapped molecules}


\author{Mudit Sinhal}
\thanks{These authors contributed equally.}
\affiliation{Department of Chemistry, University of Basel, Klingelbergstrasse 80, Basel 4056, Switzerland.}
\author{Ziv Meir}
\thanks{These authors contributed equally.}
\affiliation{Department of Chemistry, University of Basel, Klingelbergstrasse 80, Basel 4056, Switzerland.}
\author{Kaveh Najafian}
\affiliation{Department of Chemistry, University of Basel, Klingelbergstrasse 80, Basel 4056, Switzerland.}
\author{Gregor Hegi}
\affiliation{Department of Chemistry, University of Basel, Klingelbergstrasse 80, Basel 4056, Switzerland.}
\author{Stefan Willitsch}
\email[To whom correspondence should be addressed: ]{stefan.willitsch@unibas.ch}
\affiliation{Department of Chemistry, University of Basel, Klingelbergstrasse 80, Basel 4056, Switzerland.}

\date{\today}


\begin{abstract}
Trapped atoms and ions are among the best controlled quantum systems which find widespread applications in quantum information, sensing and metrology. For molecules, however, a similar degree of control is currently lacking owing to their complex energy-level structure.
Quantum-logic protocols in which atomic ions serve as probes for molecular ions are a promising route for achieving this level of control, especially with homonuclear molecules that decouple from black-body radiation. 
Here, a quantum-non-demolition protocol on single trapped \NN{} molecules is demonstrated. The spin-rovibronic state of the molecule is detected with more than 99\% fidelity and the position and strength of a spectroscopic transition in the molecule are determined, both without destroying the molecular quantum state.
The present method lays the foundations for new approaches to molecular precision spectroscopy, for state-to-state chemistry on the single-molecule level and for the implementation of molecular qubits. 
\end{abstract}

\maketitle 

The impressive advances achieved in the control of ultracold trapped atoms and ions on the quantum level are now increasingly being transferred to molecular systems. Cold, trapped molecules have been created by, e.g., binding ultracold atoms via Feshbach resonances \cite{ni08a} and photoassociation \cite{sage05a,liu18a}, molecular-beam slowing \cite{meerakker12a}, direct laser cooling \cite{barry14a,anderegg18a} and sympathetic cooling \cite{molhave00a, willitsch12a}. The trapping of the cold molecules enables experiments with long interaction times and thus paves the way for new applications such as studies of ultracold chemistry \cite{ospelkaus10b} and precision spectroscopic measurements which aim, e.g., at a precise determination of fundamental physical constants \cite{alighanbari18a} and their possible time variation \cite{schiller05a, beloy11a} as well as tests of fundamental theories which reach beyond the standard model \cite{safronova18a, demille17a}. 

The complex energy level structure and the absence of optical cycling transitions in most molecular systems constitute a major challenge in the state preparation, laser cooling, state detection and coherent manipulation of molecules. Molecular ions trapped in radiofrequency ion traps which are sympathetically cooled by simultaneously trapped atomic ions \cite{molhave00a, willitsch12a} have proven a promising route for overcoming these obstacles. Recently, their rotational cooling and state preparation has been achieved \cite{staanum10a,schneider10a,tong10a,lien14a}, precision measurements of quantum electrodynamics and fundamental constants have been performed \cite{biesheuvel16a,alighanbari18a}, the first studies of dipole-forbidden spectroscopic transitions in the mid-infrared spectral domain have been reported \cite{germann14a} and state- and energy-controlled collisions with cold atoms have been realized \cite{sikorsky18a,doerfler19a}. However, in order to reach the same exquisite level of control on the quantum level for a single molecule  which can be achieved with trapped atoms \cite{harty14a}, new methodological developments are required. In this context, the most promising route for achieving ultimate quantum control of molecular ions in trap experiments is constituted by quantum-logic protocols \cite{schmidt05a} in which a co-trapped atomic ion acts as a probe for the quantum state of a single molecular ion \cite{wolf16a,chou17a}.   

Here, a quantum-logic based quantum-non-demolition (QND) \cite{braginsky80, braginsky96, hume07a} detection of the spin-rotational-vibrational state of a single molecular nitrogen ion co-trapped with a single atomic calcium ion is demonstrated. \NN{} is a homonuclear diatomic molecule with no permanent-dipole moment rendering all ro-vibrational transitions dipole forbidden in its electronic ground state \cite{germann14a}. Therefore, \NN{} is an ideal testbed for precision spectroscopic studies \cite{kajita15a}, for tests of fundamental physics \cite{kajita14a}, for the realization of mid-IR frequency-standards and clocks \cite{schiller14a} and for the implementation of molecular qubits for quantum-information and computation applications \cite{kimble08,wehner18}.

The state-detection protocol implemented here relies on coherent motional excitation of the \Ca{}-\NN{} two-ion string \cite{meir19a,hume11a,koelemeij07a} using an optical-dipole force (ODF) which is dependent on the molecular state and arises from off-resonant dispersive molecule-light interactions. The excited motion is read out on the \Ca{} ion, thus preserving the state of the \NN{} ion. State-detection fidelities above 99\% for the ground rovibrational state of \NN{} are demonstrated, limited only by the chosen bandwidth of the detection cycle. Since the lifetime of the rovibrational levels in \NN{} is estimated to be on the order of half a year \cite{germann14a}, the prospects for reaching, and potentially exceeding, the high readout fidelities which are currently achieved with atomic ions \cite{harty14a,christensen19} are excellent.

The present scheme has immediate applications for marked improvements of the sensitivity and, therefore, precision of spectroscopic measurements on molecular ions. This is demonstrated here by introducing a type of ``force'' spectroscopy \cite{biercuk10} used to study a rovibronic component of the electronic spectrum of a single \NN{} molecule with a signal-to-noise ratio greatly exceeding previous destructive detection schemes for trapped particles. Transition properties such as the line center and the Einstein-$A$ coefficient were determined and validated against the results of previous studies which used conventional spectroscopic methods. 


\begin{figure*}
	\centering
\includegraphics[width=0.9\linewidth,trim={1.0cm 0.3cm 0cm 0cm},clip]{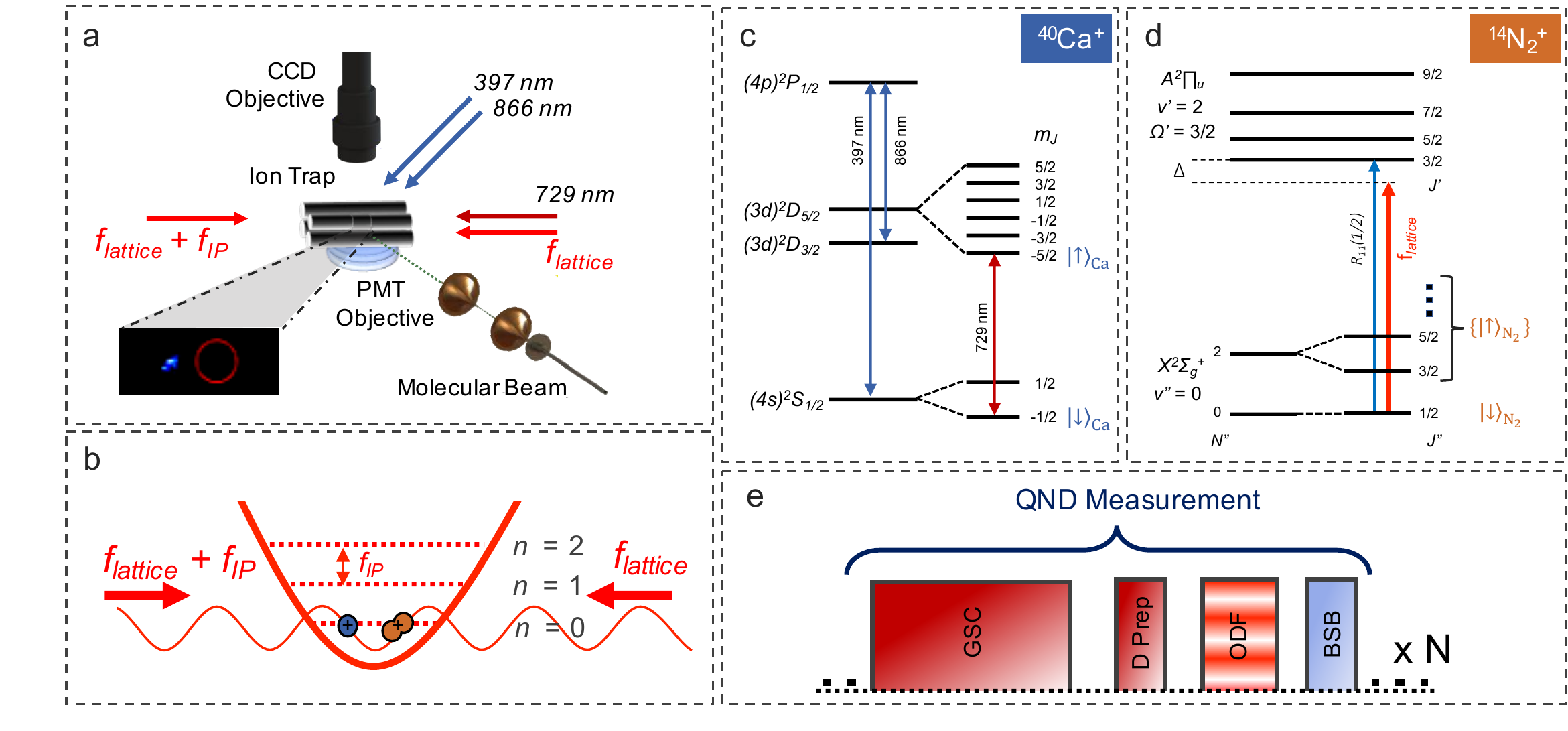}
 	\caption{\textbf{Experimental scheme.} a) Schematic representation of the experimental setup depicting the ion trap and the molecular-beam source. The fluorescence of the atomic ion (inset) is collected by a CCD camera objective. The red circle marks the position of the non-fluorescing \NN{} molecular ion in the two-ion crystal. b) Illustration of a \Ca{}-\NN{} two-ion crystal whose in-phase external motion has been cooled to the ground state of the ion trap ($n=0$) and which is overlapped with two counter-propagating laser beams forming a running one-dimensional optical lattice. c) Reduced energy-level diagram of \Ca{}. The $(4s)^2S_{1/2}\leftrightarrow (4p)^2P_{1/2}\leftrightarrow(3d)^2D_{3/2}$ closed cycling transitions (blue arrows) are used for Doppler cooling and for detection of the states \Cadown{} and \Caup{}. These states are coherently coupled by a narrow-linewidth laser beam (red arrow). d) Reduced energy-level diagram of \NN{}. The lattice laser frequency (red arrow) is detuned by $\Delta$ from the $R_{11}(1/2)$ transition (blue arrow). The state \NNdown{} is strongly coupled to the lattice while all other states \NNup{} are far detuned and hence do not couple. e) Experimental sequence of a single quantum non-demolition measurement to be repeated multiple ($N$) times to increase the fidelity of the determination of the molecular state. See text for details.}
 	\label{fig:Lattice}
\end{figure*}

Our scheme is illustrated in Fig. \ref{fig:Lattice}. A detailed description of our experimental apparatus can be found in Ref. \cite{meir19a}. Briefly, our setup consists of a molecular-beam machine coupled to a radiofrequency ion trap (Fig. \ref{fig:Lattice}a). The experiment starts by loading a small Coulomb crystal of roughly ten $^{40}$\Ca{} ions into the trap. The atomic ions are Doppler cooled to mK temperatures by scattering photons on the $(4s)^2S_{1/2} \leftrightarrow  (4p)^2P_{1/2} \leftrightarrow  (3d)^2D_{3/2}$ closed optical cycling transitions (Fig. \ref{fig:Lattice}c). A single \NN{} molecular ion is then loaded into the trap using state-selective resonance-enhanced multi-photon ionization (REMPI) from a pulsed molecular beam of neutral N$_2$ molecules \cite{tong10a,gardner19a}. This ionization scheme preferentially creates \NN{} ions in the ground electronic, vibrational and spin-rotational state, $|X^2\Sigma_g^+,v=0,N=0, J=1/2\rangle$, henceforth referred to as \NNdown{} (Fig. \ref{fig:Lattice}d). Here, $v$ denotes the vibrational, $N$ the rotational and $J$ the total-angular-momentum quantum numbers of the molecule excluding nuclear spin. The ionized molecule is sympathetically cooled by the Coulomb crystal of atomic ions \cite{willitsch12a} within a few seconds. Then, by lowering the trap depth \Ca{} ions are successively ejected from the trap while the molecular ion is retained in the trap until a \Ca{}-\NN{} two-ion crystal is obtained (see inset of Fig. \ref{fig:Lattice}a) \cite{meir19a}. 

\begin{figure}
	\centering
	\includegraphics[width=1\linewidth,trim={0.2cm 0.3cm 0.2cm 0.3cm},clip]{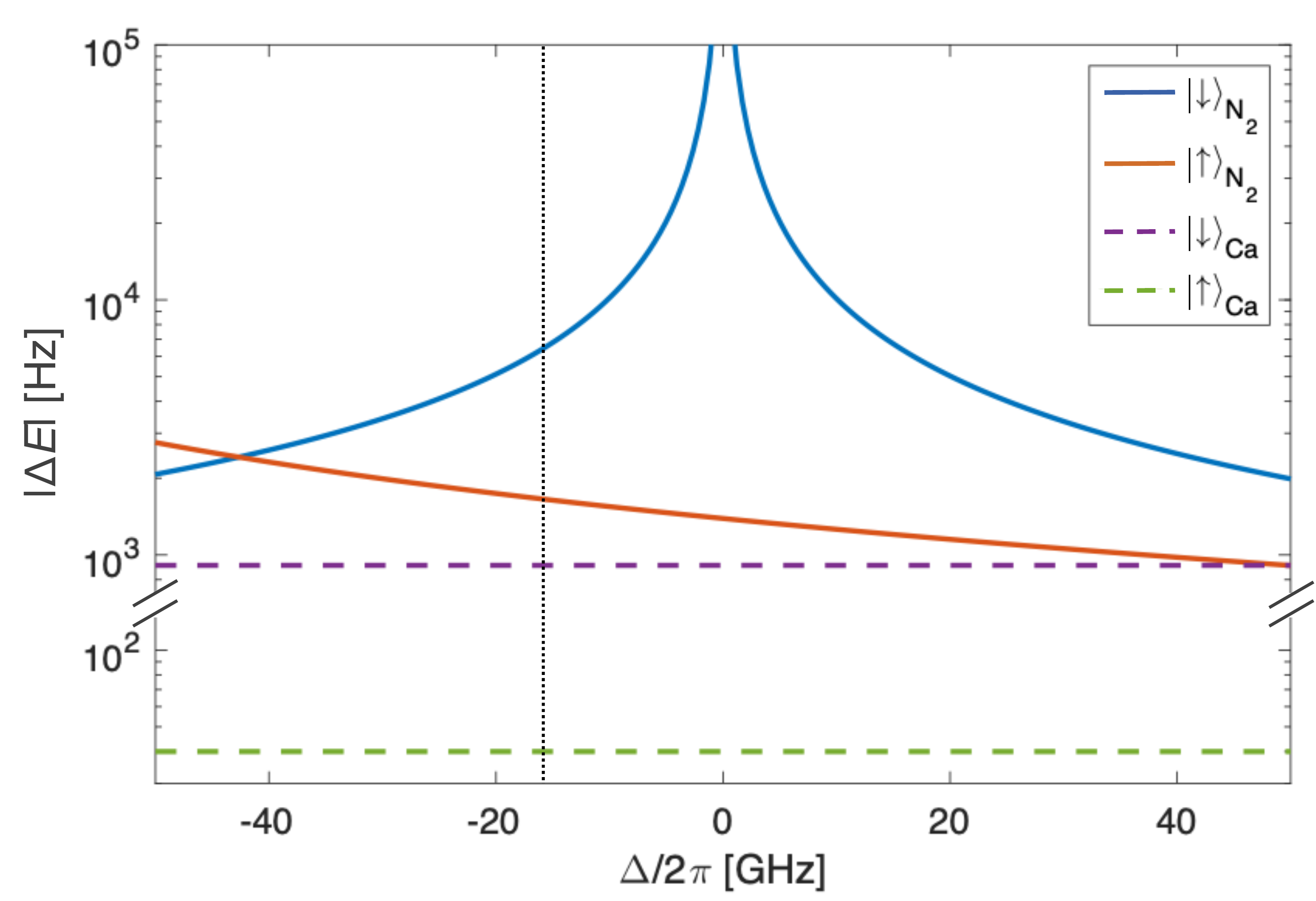}
 	\caption{\textbf{Ac-Stark shift generated on N$_2^+$ by the one-dimensional optical lattice.} Calculated magnitude of the ac-Stark shift, $|\Delta E|$, experienced by \NN{}, as a function of the laser-frequency detuning, $\Delta$, from the $A^2\Pi_{u}(v'=2) \leftarrow X^2\Sigma_{g}^+(v''=0)$, $R_{11}(1/2)$, spin-rovibronic transition \cite{herzberg91a} at $\sim$787.47 nm \cite{wu07a} for a single lattice beam of intensity 2x10$^6$ W/m$^2$. The ac-Stark shift experienced by \NN{} when in the \NNdown{} (not in the \NNdown{}) state is given by the blue (red) trace. \Ca{} in the \Caup{} (\Cadown{}) state experiences an ac-Stark shift of 40 Hz (910 Hz) indicated by the dashed green (purple) line. The black dotted line indicates the frequency detuning, $\Delta/2\pi \approx - 17$ GHz, of the optical lattice used to generate the optical dipole force which was employed for collecting the data shown in Fig. \ref{fig:N2_Flop}}
	\label{fig:N2_ODF}
\end{figure}

The experimental sequence to detect the spin-rovibrational state of the \NN{} ion is illustrated in Fig. \ref{fig:Lattice}e. First, the in-phase motional mode of the \Ca{}-\NN{} crystal is cooled to the ground state of the trap, \ket{0}, by resolved-sideband cooling (``GSC'' in Fig. \ref{fig:Lattice}e) on the atomic ion \cite{meir19a}. At the end of the cooling cycle, the \Ca{} ion is optically pumped into its $|S_{1/2},m=-1/2\rangle$ state, henceforth referred to as \Cadown{}, where $m$ denotes the magnetic quantum number. The \Ca{} ion is then shelved in the metastable $|D_{5/2},m=-5/2\rangle$ state, henceforth referred to as \Caup{}, using a $\pi$-pulse on the narrow \Caup{}$\leftarrow$\Cadown{} electric-quadruple transition followed by a $D$-purification pulse \cite{chou17a} (``D Prep'' in Fig. \ref{fig:Lattice}e). This pulse allows us to exclude experiments in which the ion is not successfully shelved to the \Caup{} state, see supplementary materials (SM) for more information. Preparing the \Ca{} ion in the \Caup{} state suppresses background signal during the state detection and thus enables an efficient determination of the state of \NN{} \cite{meir19a} (see Fig. \ref{fig:N2_ODF} and SM). For an ideal state preparation, the complete state of the two ions in the trap is then given by \NNdown{}\Caup{}\ket{0}.
 
The detection sequence is continued by applying a state-dependent optical-dipole force (ODF) to excite coherent motion \cite{meekhof96a} depending on the rotational state of the molecular ion (``ODF'' in Fig. \ref{fig:Lattice}e). In the case that the molecular ion is in the \NNdown{} state, a coherent motion of amplitude \am{} is excited such that the probability to populate a motional Fock state, \ket{n}, is given by, $P(n|\alpha$) = $|\langle n|\alpha\rangle|^2$ = $e^{-|\alpha|^2}|\alpha|^{2n}/n!$ \cite{leibfried03a}. However, if the molecular ion is in any other rotational or vibrational state, henceforth referred to as \NNup{}, a motional state \bmm{} is excited where $\{|\beta|\}\ll|\alpha|$. Here, the curly brackets are a reminder that \NNup{} refers to many states all of which result in vanishingly small amplitudes of the coherent motion, $\{|\beta|\}\ll1$. The ODF is implemented via a state-dependent ac-Stark shift generated by two counter-propagating laser beams with frequencies $f_\text{lattice}$, aligned with the crystal axis which form a one-dimensional optical lattice (Fig. \ref{fig:Lattice}b). By further detuning one of the beams by the frequency of the in-phase motional mode of the two-ion crystal, $f_{IP}\approx620$ kHz, a running lattice is generated causing a modulation of the amplitude of the ac-Stark shift which resonantly excites coherent motion of the ion crystal depending on the rotational and vibrational state of the \NN{} ion. 

Fig. \ref{fig:N2_ODF} shows the calculated ac-Stark shift of a single lattice beam as a function of its frequency for N$_2^+$ in the \NNdown{} state (blue) and the maximum ac-Stark shift experienced by the \NN{} ion when not in the \NNdown{} state (red). Details of the calculations are given in the SM. The strength of the ac-Stark shift depends on the detuning of the lattice laser beam from spectroscopic transitions in the molecule. The peak in the ac-Stark shift of the blue trace corresponds to an on-resonance condition of the $A^2\Pi_{u}(v'=2) \leftarrow X^2\Sigma_{g}^+(v''=0)$, $R_{11}(1/2)$, spin-rovibronic transition \cite{herzberg91a} originating from the \NNdown{} state (see Fig. \ref{fig:Lattice}d), where $''$($'$) denotes the lower (upper) level of the transition. By setting the lattice-laser detuning close to this resonance, the \NN{} ion experiences a much stronger ac-Stark shift leading to a large coherent motional excitation \am{} when in the \NNdown{} state as opposed to the situation were the \NN{} is not in the \NNdown{} state leading to a much weaker excitation $\{$\ket{\beta}$\}$. This ensures the state selectivity of the present scheme with respect to \NNdown{}.

The state-dependent coherent motional excitation \cite{meir19a,hume11a} maps the problem of distinguishing between the different internal states \NNdown{} and \NNup{} of the molecule to distinguishing between different excited motional states \am{} and \bmm{} of the two-ion crystal. The latter is achieved by Rabi sideband thermometry\cite{meekhof96a} on the \Ca{} ion which shares the motional state with the \NN{} ion. A blue-sideband (BSB) pulse using a narrow-linewidth laser at 729~nm drives population between \Caup{}\ket{n} $\rightarrow$ \Cadown{}\ket{n-1} states. It is followed by state-selective fluorescence on \Ca{} which projects it either to the \Caup{} ``dark'' or the \Cadown{} ``bright'' state thereby measuring the success of the BSB pulse (``BSB'' in Fig. \ref{fig:Lattice}e). The probability to project to the ``bright'' state after the BSB pulse is given by $P$(\Cadown{})$ = \sum_n P(n)~\sin^2(\Omega_nt_{729}/2)$ \cite{leibfried03a}, where $t_{729}$ is the BSB pulse time, $\Omega_n=\eta\sqrt{n}\Omega_0$ is the BSB Rabi-frequency, $\eta\approx0.1$ is the Lamb-Dicke parameter and $\Omega_0\approx (2\pi)90$~kHz is the bare Rabi frequency. The motional Fock state population distributions are given by $P(n|\alpha)$ or $P(n|\{\beta\})$ depending on the state of \NN{}. Since in general, 0$<P($\Cadown$|\{\beta\}$)$<P($\Cadown$|\alpha$)$<$1, the outcome of a single BSB pulse is insufficient to determine the motional state and hence the \NN{} internal state in a single shot. However, since the internal state of \NN{} is not changed during the measurement, the detection sequence, i.e., cooling of the two-ion string to the motional ground state, preparing \Ca{} in the \Caup{} state, exciting motion by the ODF and measuring the result by a BSB pulse, can be repeated until there is sufficient statistics to distinguish between different molecular states. The experiment therefore represents a quantum-non demolition (QND) measurement \cite{braginsky80, braginsky96,hume07a}. The situation is equivalent to distinguishing between two types of coins, $\alpha$ and $\beta$, with biased probabilities to get a heads, $h$, in a coin toss given by $0<p(h|\beta)<p(h|\alpha)<1$, by repetitively flipping one of the coins $N$ times. For $p(h|\alpha)=0.52$ and $p(h|\beta)=0.06$ a fidelity of 99.5\% can be achieved in the coin (state) determination after $N=22$ repetitive coin tosses (QND measurements) (SM).

\begin{figure*}
	\centering
	\includegraphics[width=0.85\linewidth,trim={4.3cm 0.2cm 4cm 0.6cm},clip]{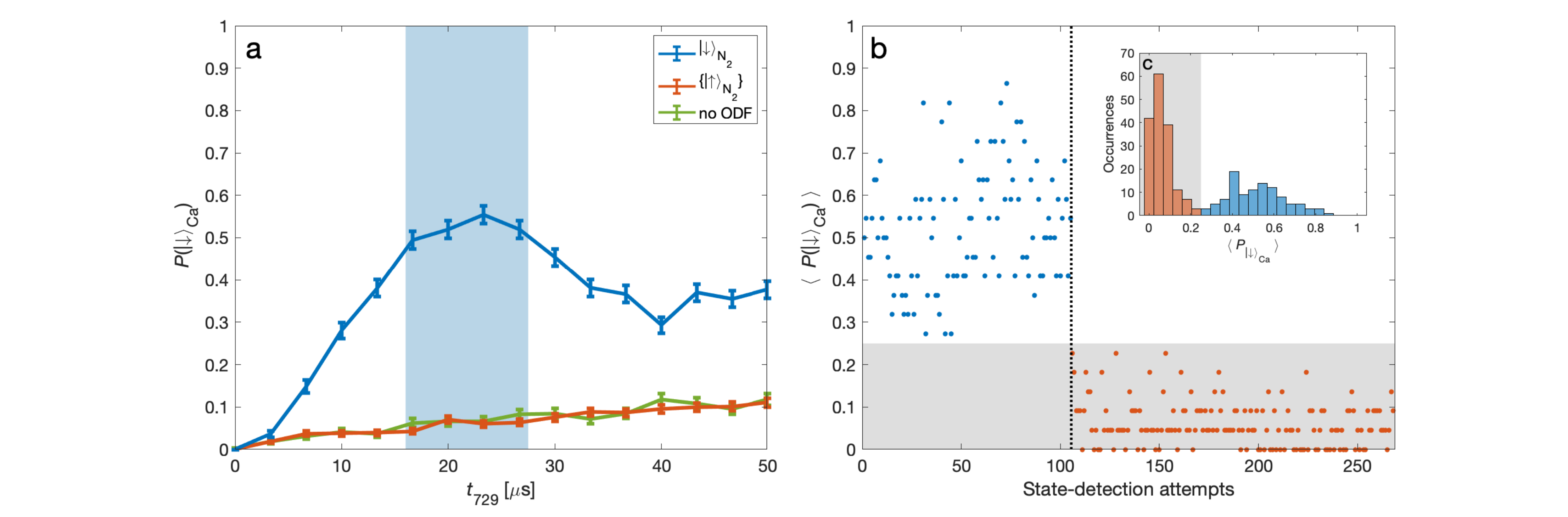}
 	\caption{\textbf{Quantum non-demolition state detection of \NN{}}. a) Blue-sideband (BSB) Rabi-oscillation signal for \NN{} in the \NNdown{} state (blue) and in one of the \NNup{} states (red). Error bars correspond to 1$\sigma$ binomial errors. Green is a background measurement of the Rabi oscillation signal without ODF beams. The light blue area indicates the region of BSB pulse lengths for which maximum state-detection contrast is achieved. b) Time trace of state-detection attempts. A single state-detection data point is composed of an average of 22 consecutive BSB-measurement results for pulse lengths indicated by the light-blue area in (a). A threshold of $P$(\Cadown{})=0.25 is used to distinguish between \NN{} in the \NNdown{} or \NNup{} states (grey area). The blue (red) dots indicate assignment of \NNdown{} (\NNup{}) states by the detection scheme. The dashed black line shows the onset of a quantum jump out of the \NNdown{} state to one of the \NNup{} states. c) Histogram of state-detection attempts. The grey area separates between \NNdown{} (blue) and \NNup{} (red) state-detection assignments.}
	\label{fig:N2_Flop}
\end{figure*}

An experimental demonstration of the present QND scheme for molecular-state detection is shown in Fig \ref{fig:N2_Flop}. Here, the ODF beams were turned on for 500 $\mu$s with a single lattice-beam intensity of $\sim2\times10^6$~W/m$^2$. The lattice lasers were detuned by $\Delta/2\pi \approx - 17$ GHz from the $R_{11}(1/2)$ transition (Fig. \ref{fig:Lattice}d). The BSB pulse time, $t_{729}$, was scanned in order to observe Rabi oscillations, shown in Fig. \ref{fig:N2_Flop}a. When the \NN{} ion was in the \NNdown{} state, a strong Rabi oscillation was observed (blue), in contrast to when the \NN{} was in one of the \NNup{} states where almost no oscillation was observed (red). The residual signal of the \NNup{} states is attributed to imperfect ground-state cooling of the two-ion crystal rather than motional excitation by the lattice beams as can be seen from a comparison to the background signal (green) obtained when the ODF beams were completely turned off.

For the parameters used in the experiment shown in Fig. \ref{fig:N2_Flop}a, the maximum contrast between the \NNdown{} and the \NNup{} signals was reached at $t_{729}\approx20$ $\mu$s. For this BSB pulse time, $P$(\Cadown{}$|\alpha$)=0.52 and $P$(\Cadown{}$|\{\beta\}$)=0.06 such that 22 QND measurements were sufficient to distinguish between the states at a confidence level of 99.5\% (SM). Such a QND determination of the \NN{} state is shown in Fig. \ref{fig:N2_Flop}b. The BSB success probability, $P$(\Cadown{}), was determined from the average of the results of 22 BSB pulses with pulse times in the range of 16.7--26.7 $\mu$s (light blue area in Fig. \ref{fig:N2_Flop}a). A threshold of $P$(\Cadown{})=0.25 was set to determine if the molecule was in the \NNdown{} (``bright molecule'') or \NNup{} (``dark molecule'') states (blue or red dots in Fig. \ref{fig:N2_Flop}b respectively). The molecular state was repeatably determined to be ``bright'' 105 times with zero false detections. Afterwards, the molecular state was repeatedly determined to be ``dark'' 163 times with zero false detections. Using Bayesian inference, the experimentally measured fidelity was 99.1(9)\% and 99.4(6)\% for the ``bright'' and ``dark'' molecule, respectively. The sudden change in the state of the molecule from ``bright'' to ``dark'' during the experiment was due to a quantum jump that was most likely caused by a state-changing collision with a background-gas molecule in our vacuum system at a pressure of $1\times10^{-10}$ mbar.

\begin{figure*}
	\centering
	\includegraphics[width=0.7\linewidth,trim={0cm 0.5cm 1cm 2.5cm},clip]{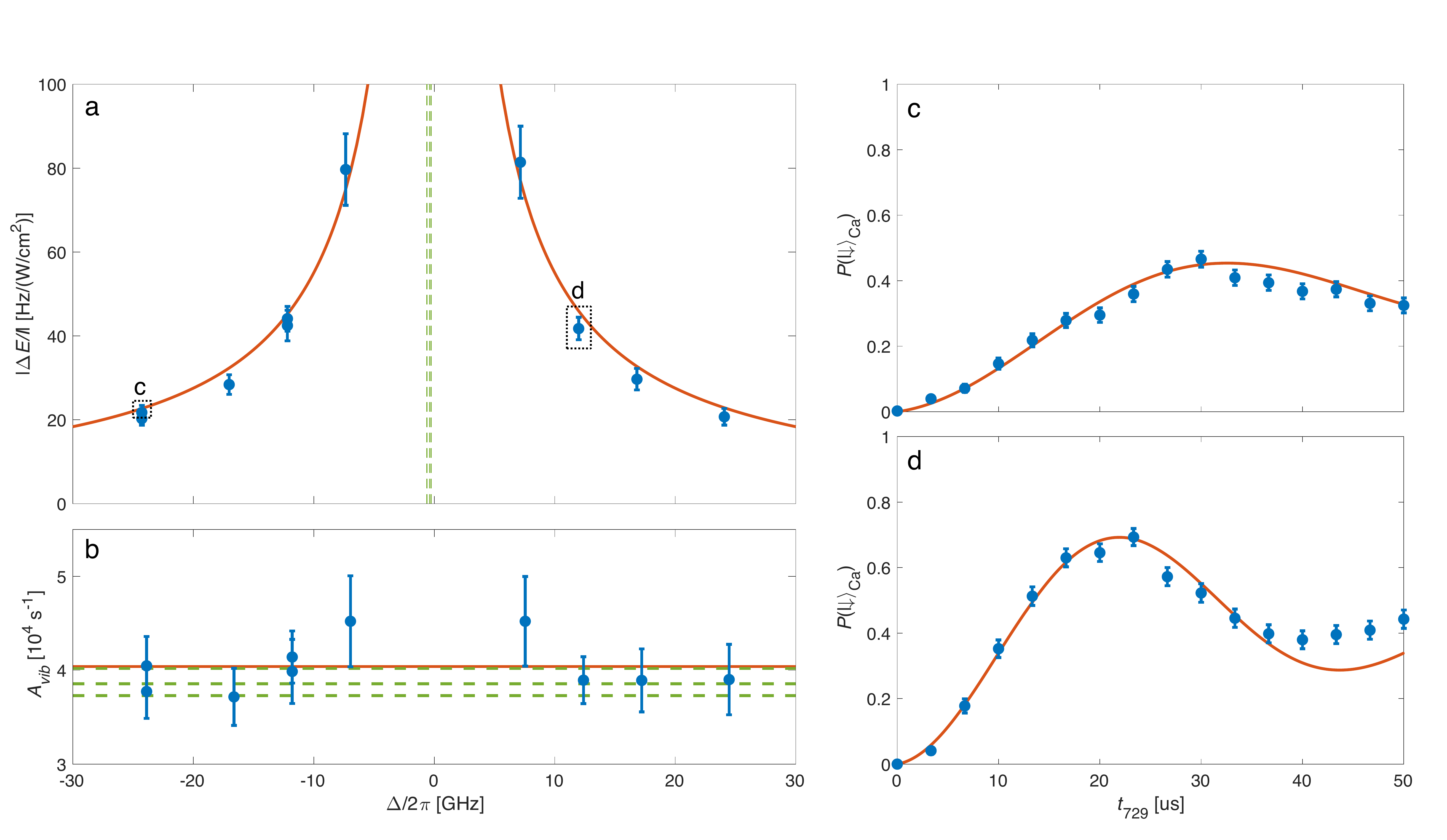}
 	\caption{\textbf{Non-destructive force spectroscopy on a single \NN{} molecule.} Spectroscopic measurement of the $A^2\Pi_{u}(v'=2)\leftarrow X^2\Sigma_{g}^+(v''=0)$, $R_{11}(1/2)$, transition in \NN{}. a) The blue data points represent the amplitude of the ac-Stark shift normalized by the lattice-laser intensity and experienced by the \NN{} ion as a function of the detuning from resonance as extracted from fits to BSB Rabi-oscillation signals (insets (c),(d)). Error bars (1$\sigma$) correspond to the uncertainty in the beam intensity and in the extraction of the ODF strength from the BSB signals. The red line is a fit to the experimental data used to extract the line center. The green dashed lines are the values of the line centers reported in the literature \cite{wu07a,bachir94a,harada94a}. b) The blue data points represent values for Einstein-$A$ coefficients of the $A^2\Pi_{u}(v'=2)\rightarrow X^2\Sigma_{g}^+(v''=0)$ vibronic transition in \NN{} extracted from the measurements in (a). The red line is the mean of all measurements. Different literature values\cite{cartwright73a,langhoff87,gilmore92} are given by the dashed green lines.}
	\label{fig:ForceSpectra}
\end{figure*}

Since the state-detection signal is proportional to the ac-Stark shift experienced by the molecule, it can be used to perform a measurement of spectroscopic transitions in the molecule. Such a spectroscopic experiment is demonstrated on the $R_{11}(1/2)$ spin-rotational component of the $A^2\Pi_{u}(v'=2)\leftarrow X^2\Sigma_{g}^+(v''=0)$ electronic-vibrational transition in \NN{} (Fig. \ref{fig:ForceSpectra}). Rabi oscillations on the BSB transition in \Ca{} resulting from the ODF acting on a molecule in the \NNdown{} state were measured for different detunings of the lattice-laser beams from this resonance. As the resonance is approached, the ac-Stark shift increases as $\sim$1/$\Delta$ leading to a larger coherent excitation, \am, of the ion crystal. The value of the ac-Stark shift was extracted from a fit to the Rabi signal (Fig. \ref{fig:ForceSpectra}c, d). The fitting function was experimentally determined by applying a well-defined force on the \Ca{} ion when the \NN{} ion was in one of the \NNup{} states and experiences no force (SM). For the chosen ODF pulse length of 500 $\mu$s, the Rabi signal is sensitive to ac-Stark shifts in the interval from 2.5 to 13 kHz. To extend the dynamic range of our measurement, the lattice-beam powers were scaled to keep the Rabi signal within the experimental sensitivity range.

Fig. \ref{fig:ForceSpectra}a depicts such a ``force'' spectrum of this transition. The experimentally measured ac-Stark shifts were fitted with a $\sim1/|f-f_0|$ ac-Stark-shift profile to determine the line center, $f_0$ = 380.7011(2)~THz, which agrees well with previous measurements \cite{wu07a,bachir94a,harada94a} using ensembles of molecules which yielded results in the range $f_0=380.7007(3)$~THz (dashed green lines in Fig. \ref{fig:ForceSpectra}a). The precision of our measurement can be enhanced by using smaller detunings, however, at the expense of an increased probability to scatter a photon at the molecule by the lattice beams and thus losing the molecular state. As an example, for the current experimental parameters of $\sim$10 GHz detuning and $\sim$10 kHz ac-Stark shift, an average of 1'000 QND state determination cycles (20'000 BSB pulses) can be expected before the molecular state is lost due to off-resonant scattering. Decreasing the detuning to 100 MHz will reduce the number of expected QND state determinations to 10 (200 BSB pulses) due to the $1/\Delta^2$ scaling of the scattering rate (compared to the ac-Stark shift scaling of $1/\Delta$). Nevertheless, these measurements are possible with efficient and reliable replenishment of molecular ions in the trap. For such close detunings, our method is expected to be sensitive to the hyperfine structure of the transition  \cite{bruna08} which is not resolved here and has, to our knowledge, not been studied experimentally yet. In this experiment, the absolute accuracy of the wavemeter used to evaluate the lattice-laser frequency was estimated to be better than 50 MHz by a repetitive measurements of the $P_{3/2} \leftarrow D_{5/2}$ spectroscopic transition in \Ca{} during the experiment.

The vibronic Einstein-$A$ coefficient, $A_{vib}$, of the $A^2\Pi_{u}(v'=2) \rightarrow X^2\Sigma_{g}^+(v''=0)$ transition was extracted from the ac-Stark-shift measurements as shown in Fig. \ref{fig:ForceSpectra}b. For each ac-Stark-shift determination, $\Delta E(\Delta)$ (Fig. \ref{fig:ForceSpectra}a), a corresponding value  $A_{vib}(\Delta)$ was calculated. The mean value of $A_{vib}$ = 4.03(11) $\times$ 10$^4$ s$^{-1}$ is in good agreement with previous results in the range $A_\text{vib}=3.87(14) \times 10^4$ s$^{-1}$ \cite{cartwright73a,langhoff87,gilmore92} (Fig. \ref{fig:ForceSpectra}b dashed green lines). The two data points with the smallest detuning in Fig. \ref{fig:ForceSpectra}b seem to slightly deviate from the other points. This might be due to the unresolved hyperfine structure of the transition which becomes non-negligible at close detunings. Nevertheless, all points were included in the determination of $A_{vib}$.


To summarize, a QND detection of the internal quantum state of a single molecule with more than 99\% fidelity has been demonstrated. The demonstrated fidelity of the state detection is not limited by the state lifetime and off-resonant scattering as in atomic ions \cite{harty14a,christensen19} and hence it can be increased even further at the expense of slower data acquisition rate. Based on this detection scheme, a new approach for measuring spectroscopic line positions and transition strengths in molecules using ``force'' spectroscopy has been realized. 

The present approach can be compared with the pioneering experiment of Wolf et al. \cite{wolf16a} in which a motional qubit was implemented to detect the internal states of polar MgH$^+$ molecules. The present scheme allows a simpler approach for state-detection by exciting coherent motion which also readily enables the extraction of accurate values for spectroscopic quantities like transition strengths. Here, a new set of tools was developed for the state preparation and quantum manipulation of apolar \NN{} molecules which do not couple to the black-body-radiation field. Although the immunity to black-body radiation imposes additional technical challenges, it makes apolar species like \NN{} attractive systems as demonstrated here by the excellent state-detection fidelity.

The present QND scheme is universally applicable to both polar and apolar molecular ions. It represents a highly sensitive method to repeatedly and non-destructively read out the quantum state of a molecule and thus introduces a molecular counterpart to the state-dependent fluorescence on closed-cycling transitions which forms the basis of sensitive readout schemes in atomic systems \cite{harty14a,christensen19}. It enables state-selected and coherent experiments with single trapped molecules with duty cycles several orders of magnitude higher than previous destructive state-detection schemes \cite{staanum10a,germann14a}. It thus lays the foundations for vast improvements in the sensitivity and, therefore, precision of spectroscopic experiments on molecular ions, as discussed in Ref. \cite{meir19a}. The possibility for efficient, non-destructive state readout also lays the foundation for the application of molecular ions in quantum-information and coherent-control experiments as are currently being performed with great success using atomic ions \cite{harty14a}. In this context, the potentially very long lifetimes and coherence times of molecular states may offer new possibilities for, e.g., realising quantum memories. As another application, the present scheme also enables studies of cold collisions and chemical reactions between ions and neutrals with state control on the single-molecule level and, therefore, offers prospects for the exploration of molecular collisions and chemical reaction mechanisms in unprecedented detail. Finally, our approach can also be employed not only to {\sl detect}, but also to {\sl prepare} the quantum state of a single molecule through a projective measurement down to the Zeeman level \cite{meir19a} opposed to previous schemes \cite{staanum10a,tong10a} which could prepare only the rovibronic level of the molecule. The present scheme thus also represents a key element in the methodological toolbox of the upcoming field of molecular quantum technologies and the realization of molecular qubits encoded in the rovibrational spectrum.

We acknowledge funding from the Swiss National Science Foundation as part of the National Centre of Competence in Research, Quantum Science and Technology (NCCR-QSIT), grant nr. CRSII5\_183579 and the University of Basel. We thank Dr. Anatoly Johnson, Philipp Kn\"opfel, Grischa Martin and Georg Holderried for technical support.



\begin{thebibliography}{55}%
\makeatletter
\providecommand \@ifxundefined [1]{%
 \@ifx{#1\undefined}
}%
\providecommand \@ifnum [1]{%
 \ifnum #1\expandafter \@firstoftwo
 \else \expandafter \@secondoftwo
 \fi
}%
\providecommand \@ifx [1]{%
 \ifx #1\expandafter \@firstoftwo
 \else \expandafter \@secondoftwo
 \fi
}%
\providecommand \natexlab [1]{#1}%
\providecommand \enquote  [1]{``#1''}%
\providecommand \bibnamefont  [1]{#1}%
\providecommand \bibfnamefont [1]{#1}%
\providecommand \citenamefont [1]{#1}%
\providecommand \href@noop [0]{\@secondoftwo}%
\providecommand \href [0]{\begingroup \@sanitize@url \@href}%
\providecommand \@href[1]{\@@startlink{#1}\@@href}%
\providecommand \@@href[1]{\endgroup#1\@@endlink}%
\providecommand \@sanitize@url [0]{\catcode `\\12\catcode `\$12\catcode
  `\&12\catcode `\#12\catcode `\^12\catcode `\_12\catcode `\%12\relax}%
\providecommand \@@startlink[1]{}%
\providecommand \@@endlink[0]{}%
\providecommand \url  [0]{\begingroup\@sanitize@url \@url }%
\providecommand \@url [1]{\endgroup\@href {#1}{\urlprefix }}%
\providecommand \urlprefix  [0]{URL }%
\providecommand \Eprint [0]{\href }%
\providecommand \doibase [0]{http://dx.doi.org/}%
\providecommand \selectlanguage [0]{\@gobble}%
\providecommand \bibinfo  [0]{\@secondoftwo}%
\providecommand \bibfield  [0]{\@secondoftwo}%
\providecommand \translation [1]{[#1]}%
\providecommand \BibitemOpen [0]{}%
\providecommand \bibitemStop [0]{}%
\providecommand \bibitemNoStop [0]{.\EOS\space}%
\providecommand \EOS [0]{\spacefactor3000\relax}%
\providecommand \BibitemShut  [1]{\csname bibitem#1\endcsname}%
\let\auto@bib@innerbib\@empty
\bibitem [{\citenamefont {Ni}\ \emph {et~al.}(2008)\citenamefont {Ni},
  \citenamefont {Ospelkaus}, \citenamefont {{\mbox{de Miranda}}}, \citenamefont
  {\mbox{Pe'er}}, \citenamefont {Neyenhuis}, \citenamefont {Zirbel},
  \citenamefont {Kotochigova}, \citenamefont {Julienne}, \citenamefont {Jin},\
  and\ \citenamefont {Ye}}]{ni08a}%
  \BibitemOpen
  \bibfield  {author} {\bibinfo {author} {\bibfnamefont {K.-K.}\ \bibnamefont
  {Ni}}, \bibinfo {author} {\bibfnamefont {S.}~\bibnamefont {Ospelkaus}},
  \bibinfo {author} {\bibfnamefont {M.~H.~G.}\ \bibnamefont {{\mbox{de
  Miranda}}}}, \bibinfo {author} {\bibfnamefont {A.}~\bibnamefont
  {\mbox{Pe'er}}}, \bibinfo {author} {\bibfnamefont {B.}~\bibnamefont
  {Neyenhuis}}, \bibinfo {author} {\bibfnamefont {J.~J.}\ \bibnamefont
  {Zirbel}}, \bibinfo {author} {\bibfnamefont {S.}~\bibnamefont {Kotochigova}},
  \bibinfo {author} {\bibfnamefont {P.~S.}\ \bibnamefont {Julienne}}, \bibinfo
  {author} {\bibfnamefont {D.~S.}\ \bibnamefont {Jin}}, \ and\ \bibinfo
  {author} {\bibfnamefont {J.}~\bibnamefont {Ye}},\ }\href@noop {} {\bibfield
  {journal} {\bibinfo  {journal} {Science}\ }\textbf {\bibinfo {volume}
  {322}},\ \bibinfo {pages} {231} (\bibinfo {year} {2008})}\BibitemShut
  {NoStop}%
\bibitem [{\citenamefont {Sage}\ \emph {et~al.}(2005)\citenamefont {Sage},
  \citenamefont {Sainis}, \citenamefont {Bergeman},\ and\ \citenamefont
  {DeMille}}]{sage05a}%
  \BibitemOpen
  \bibfield  {author} {\bibinfo {author} {\bibfnamefont {J.~M.}\ \bibnamefont
  {Sage}}, \bibinfo {author} {\bibfnamefont {S.}~\bibnamefont {Sainis}},
  \bibinfo {author} {\bibfnamefont {T.}~\bibnamefont {Bergeman}}, \ and\
  \bibinfo {author} {\bibfnamefont {D.}~\bibnamefont {DeMille}},\ }\href@noop
  {} {\bibfield  {journal} {\bibinfo  {journal} {Phys. Rev. Lett.}\ }\textbf
  {\bibinfo {volume} {94}},\ \bibinfo {pages} {203001} (\bibinfo {year}
  {2005})}\BibitemShut {NoStop}%
\bibitem [{\citenamefont {Liu}\ \emph {et~al.}(2018)\citenamefont {Liu},
  \citenamefont {Hood}, \citenamefont {Yu}, \citenamefont {Zhang},
  \citenamefont {Hutzler}, \citenamefont {Rosenband},\ and\ \citenamefont
  {Ni}}]{liu18a}%
  \BibitemOpen
  \bibfield  {author} {\bibinfo {author} {\bibfnamefont {L.~R.}\ \bibnamefont
  {Liu}}, \bibinfo {author} {\bibfnamefont {J.~D.}\ \bibnamefont {Hood}},
  \bibinfo {author} {\bibfnamefont {Y.}~\bibnamefont {Yu}}, \bibinfo {author}
  {\bibfnamefont {J.~T.}\ \bibnamefont {Zhang}}, \bibinfo {author}
  {\bibfnamefont {N.~R.}\ \bibnamefont {Hutzler}}, \bibinfo {author}
  {\bibfnamefont {T.}~\bibnamefont {Rosenband}}, \ and\ \bibinfo {author}
  {\bibfnamefont {K.-K.}\ \bibnamefont {Ni}},\ }\href@noop {} {\bibfield
  {journal} {\bibinfo  {journal} {Science}\ }\textbf {\bibinfo {volume}
  {360}},\ \bibinfo {pages} {900} (\bibinfo {year} {2018})}\BibitemShut
  {NoStop}%
\bibitem [{\citenamefont {{\mbox{van de} Meerakker}}\ \emph
  {et~al.}(2012)\citenamefont {{\mbox{van de} Meerakker}}, \citenamefont
  {Bethlem}, \citenamefont {Vanhaecke},\ and\ \citenamefont
  {Meijer}}]{meerakker12a}%
  \BibitemOpen
  \bibfield  {author} {\bibinfo {author} {\bibfnamefont {S.~Y.~T.}\
  \bibnamefont {{\mbox{van de} Meerakker}}}, \bibinfo {author} {\bibfnamefont
  {H.~L.}\ \bibnamefont {Bethlem}}, \bibinfo {author} {\bibfnamefont
  {N.}~\bibnamefont {Vanhaecke}}, \ and\ \bibinfo {author} {\bibfnamefont
  {G.}~\bibnamefont {Meijer}},\ }\href@noop {} {\bibfield  {journal} {\bibinfo
  {journal} {Chem. Rev.}\ }\textbf {\bibinfo {volume} {112}},\ \bibinfo {pages}
  {4828} (\bibinfo {year} {2012})}\BibitemShut {NoStop}%
\bibitem [{\citenamefont {Barry}\ \emph {et~al.}(2014)\citenamefont {Barry},
  \citenamefont {McCarron}, \citenamefont {Norrgard}, \citenamefont
  {Steinecker},\ and\ \citenamefont {DeMille}}]{barry14a}%
  \BibitemOpen
  \bibfield  {author} {\bibinfo {author} {\bibfnamefont {J.}~\bibnamefont
  {Barry}}, \bibinfo {author} {\bibfnamefont {D.}~\bibnamefont {McCarron}},
  \bibinfo {author} {\bibfnamefont {E.}~\bibnamefont {Norrgard}}, \bibinfo
  {author} {\bibfnamefont {M.}~\bibnamefont {Steinecker}}, \ and\ \bibinfo
  {author} {\bibfnamefont {D.}~\bibnamefont {DeMille}},\ }\href@noop {}
  {\bibfield  {journal} {\bibinfo  {journal} {Nature}\ }\textbf {\bibinfo
  {volume} {512}},\ \bibinfo {pages} {286} (\bibinfo {year}
  {2014})}\BibitemShut {NoStop}%
\bibitem [{\citenamefont {Anderegg}\ \emph {et~al.}(2018)\citenamefont
  {Anderegg}, \citenamefont {Augenbraun}, \citenamefont {Bao}, \citenamefont
  {Burchesky}, \citenamefont {Cheuk}, \citenamefont {Ketterle},\ and\
  \citenamefont {Doyle}}]{anderegg18a}%
  \BibitemOpen
  \bibfield  {author} {\bibinfo {author} {\bibfnamefont {L.}~\bibnamefont
  {Anderegg}}, \bibinfo {author} {\bibfnamefont {B.~L.}\ \bibnamefont
  {Augenbraun}}, \bibinfo {author} {\bibfnamefont {Y.}~\bibnamefont {Bao}},
  \bibinfo {author} {\bibfnamefont {S.}~\bibnamefont {Burchesky}}, \bibinfo
  {author} {\bibfnamefont {L.~W.}\ \bibnamefont {Cheuk}}, \bibinfo {author}
  {\bibfnamefont {W.}~\bibnamefont {Ketterle}}, \ and\ \bibinfo {author}
  {\bibfnamefont {J.~M.}\ \bibnamefont {Doyle}},\ }\href@noop {} {\bibfield
  {journal} {\bibinfo  {journal} {Nat. Phys.}\ }\textbf {\bibinfo {volume}
  {14}},\ \bibinfo {pages} {890} (\bibinfo {year} {2018})}\BibitemShut
  {NoStop}%
\bibitem [{\citenamefont {M{\o}lhave}\ and\ \citenamefont
  {Drewsen}(2000)}]{molhave00a}%
  \BibitemOpen
  \bibfield  {author} {\bibinfo {author} {\bibfnamefont {K.}~\bibnamefont
  {M{\o}lhave}}\ and\ \bibinfo {author} {\bibfnamefont {M.}~\bibnamefont
  {Drewsen}},\ }\href@noop {} {\bibfield  {journal} {\bibinfo  {journal}
  {{Phys. Rev. A}}\ }\textbf {\bibinfo {volume} {62}},\ \bibinfo {pages}
  {011401} (\bibinfo {year} {2000})}\BibitemShut {NoStop}%
\bibitem [{\citenamefont {Willitsch}(2012)}]{willitsch12a}%
  \BibitemOpen
  \bibfield  {author} {\bibinfo {author} {\bibfnamefont {S.}~\bibnamefont
  {Willitsch}},\ }\href@noop {} {\bibfield  {journal} {\bibinfo  {journal}
  {Int. Rev. Phys. Chem.}\ }\textbf {\bibinfo {volume} {31}},\ \bibinfo {pages}
  {175} (\bibinfo {year} {2012})}\BibitemShut {NoStop}%
\bibitem [{\citenamefont {Ospelkaus}\ \emph {et~al.}(2010)\citenamefont
  {Ospelkaus}, \citenamefont {Ni}, \citenamefont {Wang}, \citenamefont
  {{\mbox{de Miranda}}}, \citenamefont {Neyenhuis}, \citenamefont
  {Qu{\'e}m{\'e}ner}, \citenamefont {Julienne}, \citenamefont {Bohn},
  \citenamefont {Jin},\ and\ \citenamefont {Ye}}]{ospelkaus10b}%
  \BibitemOpen
  \bibfield  {author} {\bibinfo {author} {\bibfnamefont {S.}~\bibnamefont
  {Ospelkaus}}, \bibinfo {author} {\bibfnamefont {K.-K.}\ \bibnamefont {Ni}},
  \bibinfo {author} {\bibfnamefont {D.}~\bibnamefont {Wang}}, \bibinfo {author}
  {\bibfnamefont {M.~H.~G.}\ \bibnamefont {{\mbox{de Miranda}}}}, \bibinfo
  {author} {\bibfnamefont {B.}~\bibnamefont {Neyenhuis}}, \bibinfo {author}
  {\bibfnamefont {G.}~\bibnamefont {Qu{\'e}m{\'e}ner}}, \bibinfo {author}
  {\bibfnamefont {P.~S.}\ \bibnamefont {Julienne}}, \bibinfo {author}
  {\bibfnamefont {J.~L.}\ \bibnamefont {Bohn}}, \bibinfo {author}
  {\bibfnamefont {D.~S.}\ \bibnamefont {Jin}}, \ and\ \bibinfo {author}
  {\bibfnamefont {J.}~\bibnamefont {Ye}},\ }\href@noop {} {\bibfield  {journal}
  {\bibinfo  {journal} {Science}\ }\textbf {\bibinfo {volume} {327}},\ \bibinfo
  {pages} {853} (\bibinfo {year} {2010})}\BibitemShut {NoStop}%
\bibitem [{\citenamefont {Alighanbari}\ \emph {et~al.}(2018)\citenamefont
  {Alighanbari}, \citenamefont {Hansen}, \citenamefont {Korobov},\ and\
  \citenamefont {Schiller}}]{alighanbari18a}%
  \BibitemOpen
  \bibfield  {author} {\bibinfo {author} {\bibfnamefont {S.}~\bibnamefont
  {Alighanbari}}, \bibinfo {author} {\bibfnamefont {M.~G.}\ \bibnamefont
  {Hansen}}, \bibinfo {author} {\bibfnamefont {V.~I.}\ \bibnamefont {Korobov}},
  \ and\ \bibinfo {author} {\bibfnamefont {S.}~\bibnamefont {Schiller}},\
  }\href@noop {} {\bibfield  {journal} {\bibinfo  {journal} {Nat. Phys.}\
  }\textbf {\bibinfo {volume} {14}},\ \bibinfo {pages} {555} (\bibinfo {year}
  {2018})}\BibitemShut {NoStop}%
\bibitem [{\citenamefont {Schiller}\ and\ \citenamefont
  {Korobov}(2005)}]{schiller05a}%
  \BibitemOpen
  \bibfield  {author} {\bibinfo {author} {\bibfnamefont {S.}~\bibnamefont
  {Schiller}}\ and\ \bibinfo {author} {\bibfnamefont {V.}~\bibnamefont
  {Korobov}},\ }\href {2505} {\bibfield  {journal} {\bibinfo  {journal} {{Phys.
  Rev. A}}\ }\textbf {\bibinfo {volume} {71}},\ \bibinfo {pages} {032505}
  (\bibinfo {year} {2005})}\BibitemShut {NoStop}%
\bibitem [{\citenamefont {Beloy}\ \emph {et~al.}(2011)\citenamefont {Beloy},
  \citenamefont {Kozlov}, \citenamefont {Borschevsky}, \citenamefont {Hauser},
  \citenamefont {Flambaum},\ and\ \citenamefont {Schwerdtfeger}}]{beloy11a}%
  \BibitemOpen
  \bibfield  {author} {\bibinfo {author} {\bibfnamefont {K.}~\bibnamefont
  {Beloy}}, \bibinfo {author} {\bibfnamefont {M.~G.}\ \bibnamefont {Kozlov}},
  \bibinfo {author} {\bibfnamefont {A.}~\bibnamefont {Borschevsky}}, \bibinfo
  {author} {\bibfnamefont {A.~W.}\ \bibnamefont {Hauser}}, \bibinfo {author}
  {\bibfnamefont {V.~V.}\ \bibnamefont {Flambaum}}, \ and\ \bibinfo {author}
  {\bibfnamefont {P.}~\bibnamefont {Schwerdtfeger}},\ }\href@noop {} {\bibfield
   {journal} {\bibinfo  {journal} {Phys. Rev. A}\ }\textbf {\bibinfo {volume}
  {83}},\ \bibinfo {pages} {062514} (\bibinfo {year} {2011})}\BibitemShut
  {NoStop}%
\bibitem [{\citenamefont {Safronova}\ \emph {et~al.}(2018)\citenamefont
  {Safronova}, \citenamefont {Budker}, \citenamefont {DeMille}, \citenamefont
  {Kimball}, \citenamefont {Derevianko},\ and\ \citenamefont
  {Clark}}]{safronova18a}%
  \BibitemOpen
  \bibfield  {author} {\bibinfo {author} {\bibfnamefont {M.~S.}\ \bibnamefont
  {Safronova}}, \bibinfo {author} {\bibfnamefont {D.}~\bibnamefont {Budker}},
  \bibinfo {author} {\bibfnamefont {D.}~\bibnamefont {DeMille}}, \bibinfo
  {author} {\bibfnamefont {D.~F.~J.}\ \bibnamefont {Kimball}}, \bibinfo
  {author} {\bibfnamefont {A.}~\bibnamefont {Derevianko}}, \ and\ \bibinfo
  {author} {\bibfnamefont {C.~W.}\ \bibnamefont {Clark}},\ }\href@noop {}
  {\bibfield  {journal} {\bibinfo  {journal} {Rev. Mod. Phys.}\ }\textbf
  {\bibinfo {volume} {90}},\ \bibinfo {pages} {025008} (\bibinfo {year}
  {2018})}\BibitemShut {NoStop}%
\bibitem [{\citenamefont {DeMille}\ \emph {et~al.}(2017)\citenamefont
  {DeMille}, \citenamefont {Doyle},\ and\ \citenamefont
  {Sushkov}}]{demille17a}%
  \BibitemOpen
  \bibfield  {author} {\bibinfo {author} {\bibfnamefont {D.}~\bibnamefont
  {DeMille}}, \bibinfo {author} {\bibfnamefont {J.~M.}\ \bibnamefont {Doyle}},
  \ and\ \bibinfo {author} {\bibfnamefont {A.~O.}\ \bibnamefont {Sushkov}},\
  }\href@noop {} {\bibfield  {journal} {\bibinfo  {journal} {Science}\ }\textbf
  {\bibinfo {volume} {357}},\ \bibinfo {pages} {990} (\bibinfo {year}
  {2017})}\BibitemShut {NoStop}%
\bibitem [{\citenamefont {Staanum}\ \emph {et~al.}(2010)\citenamefont
  {Staanum}, \citenamefont {H{\o}jbjerre}, \citenamefont {Skyt}, \citenamefont
  {Hansen},\ and\ \citenamefont {Drewsen}}]{staanum10a}%
  \BibitemOpen
  \bibfield  {author} {\bibinfo {author} {\bibfnamefont {P.~F.}\ \bibnamefont
  {Staanum}}, \bibinfo {author} {\bibfnamefont {K.}~\bibnamefont
  {H{\o}jbjerre}}, \bibinfo {author} {\bibfnamefont {P.~S.}\ \bibnamefont
  {Skyt}}, \bibinfo {author} {\bibfnamefont {A.~K.}\ \bibnamefont {Hansen}}, \
  and\ \bibinfo {author} {\bibfnamefont {M.}~\bibnamefont {Drewsen}},\
  }\href@noop {} {\bibfield  {journal} {\bibinfo  {journal} {Nat. Phys.}\
  }\textbf {\bibinfo {volume} {6}},\ \bibinfo {pages} {271} (\bibinfo {year}
  {2010})}\BibitemShut {NoStop}%
\bibitem [{\citenamefont {Schneider}\ \emph {et~al.}(2010)\citenamefont
  {Schneider}, \citenamefont {Roth}, \citenamefont {Duncker}, \citenamefont
  {Ernsting},\ and\ \citenamefont {Schiller}}]{schneider10a}%
  \BibitemOpen
  \bibfield  {author} {\bibinfo {author} {\bibfnamefont {T.}~\bibnamefont
  {Schneider}}, \bibinfo {author} {\bibfnamefont {B.}~\bibnamefont {Roth}},
  \bibinfo {author} {\bibfnamefont {H.}~\bibnamefont {Duncker}}, \bibinfo
  {author} {\bibfnamefont {I.}~\bibnamefont {Ernsting}}, \ and\ \bibinfo
  {author} {\bibfnamefont {S.}~\bibnamefont {Schiller}},\ }\href@noop {}
  {\bibfield  {journal} {\bibinfo  {journal} {Nat. Phys.}\ }\textbf {\bibinfo
  {volume} {6}},\ \bibinfo {pages} {275} (\bibinfo {year} {2010})}\BibitemShut
  {NoStop}%
\bibitem [{\citenamefont {Tong}\ \emph {et~al.}(2010)\citenamefont {Tong},
  \citenamefont {Winney},\ and\ \citenamefont {Willitsch}}]{tong10a}%
  \BibitemOpen
  \bibfield  {author} {\bibinfo {author} {\bibfnamefont {X.}~\bibnamefont
  {Tong}}, \bibinfo {author} {\bibfnamefont {A.~H.}\ \bibnamefont {Winney}}, \
  and\ \bibinfo {author} {\bibfnamefont {S.}~\bibnamefont {Willitsch}},\
  }\href@noop {} {\bibfield  {journal} {\bibinfo  {journal} {{Phys. Rev.
  Lett.}}\ }\textbf {\bibinfo {volume} {105}},\ \bibinfo {pages} {143001}
  (\bibinfo {year} {2010})}\BibitemShut {NoStop}%
\bibitem [{\citenamefont {Lien}\ \emph {et~al.}(2014)\citenamefont {Lien},
  \citenamefont {Seck}, \citenamefont {Lin}, \citenamefont {Nguyen},
  \citenamefont {Tabor},\ and\ \citenamefont {Odom}}]{lien14a}%
  \BibitemOpen
  \bibfield  {author} {\bibinfo {author} {\bibfnamefont {C.-Y.}\ \bibnamefont
  {Lien}}, \bibinfo {author} {\bibfnamefont {C.~M.}\ \bibnamefont {Seck}},
  \bibinfo {author} {\bibfnamefont {Y.-W.}\ \bibnamefont {Lin}}, \bibinfo
  {author} {\bibfnamefont {J.~H.~V.}\ \bibnamefont {Nguyen}}, \bibinfo {author}
  {\bibfnamefont {D.~A.}\ \bibnamefont {Tabor}}, \ and\ \bibinfo {author}
  {\bibfnamefont {B.~C.}\ \bibnamefont {Odom}},\ }\href@noop {} {\bibfield
  {journal} {\bibinfo  {journal} {Nat. Commun.}\ }\textbf {\bibinfo {volume}
  {5}},\ \bibinfo {pages} {4783} (\bibinfo {year} {2014})}\BibitemShut
  {NoStop}%
\bibitem [{\citenamefont {Biesheuvel}\ \emph {et~al.}(2016)\citenamefont
  {Biesheuvel}, \citenamefont {Karr}, \citenamefont {Hilico}, \citenamefont
  {Eikema}, \citenamefont {Ubachs},\ and\ \citenamefont
  {Koelemeij}}]{biesheuvel16a}%
  \BibitemOpen
  \bibfield  {author} {\bibinfo {author} {\bibfnamefont {J.}~\bibnamefont
  {Biesheuvel}}, \bibinfo {author} {\bibfnamefont {J.~P.}\ \bibnamefont
  {Karr}}, \bibinfo {author} {\bibfnamefont {L.}~\bibnamefont {Hilico}},
  \bibinfo {author} {\bibfnamefont {K.~S.~E.}\ \bibnamefont {Eikema}}, \bibinfo
  {author} {\bibfnamefont {W.}~\bibnamefont {Ubachs}}, \ and\ \bibinfo {author}
  {\bibfnamefont {J.~C.~J.}\ \bibnamefont {Koelemeij}},\ }\href@noop {}
  {\bibfield  {journal} {\bibinfo  {journal} {Nat. Commun.}\ }\textbf {\bibinfo
  {volume} {7}},\ \bibinfo {pages} {10385} (\bibinfo {year}
  {2016})}\BibitemShut {NoStop}%
\bibitem [{\citenamefont {Germann}\ \emph {et~al.}(2014)\citenamefont
  {Germann}, \citenamefont {Tong},\ and\ \citenamefont
  {Willitsch}}]{germann14a}%
  \BibitemOpen
  \bibfield  {author} {\bibinfo {author} {\bibfnamefont {M.}~\bibnamefont
  {Germann}}, \bibinfo {author} {\bibfnamefont {X.}~\bibnamefont {Tong}}, \
  and\ \bibinfo {author} {\bibfnamefont {S.}~\bibnamefont {Willitsch}},\
  }\href@noop {} {\bibfield  {journal} {\bibinfo  {journal} {Nat. Phys.}\
  }\textbf {\bibinfo {volume} {10}},\ \bibinfo {pages} {820} (\bibinfo {year}
  {2014})}\BibitemShut {NoStop}%
\bibitem [{\citenamefont {Sikorsky}\ \emph {et~al.}(2018)\citenamefont
  {Sikorsky}, \citenamefont {Meir}, \citenamefont {Ben-shlomi}, \citenamefont
  {Akerman},\ and\ \citenamefont {Ozeri}}]{sikorsky18a}%
  \BibitemOpen
  \bibfield  {author} {\bibinfo {author} {\bibfnamefont {T.}~\bibnamefont
  {Sikorsky}}, \bibinfo {author} {\bibfnamefont {Z.}~\bibnamefont {Meir}},
  \bibinfo {author} {\bibfnamefont {R.}~\bibnamefont {Ben-shlomi}}, \bibinfo
  {author} {\bibfnamefont {N.}~\bibnamefont {Akerman}}, \ and\ \bibinfo
  {author} {\bibfnamefont {R.}~\bibnamefont {Ozeri}},\ }\href@noop {}
  {\bibfield  {journal} {\bibinfo  {journal} {Nat. Commun.}\ }\textbf {\bibinfo
  {volume} {9}},\ \bibinfo {pages} {920} (\bibinfo {year} {2018})}\BibitemShut
  {NoStop}%
\bibitem [{\citenamefont {D{\"o}rfler}\ \emph {et~al.}(2019)\citenamefont
  {D{\"o}rfler}, \citenamefont {Eberle}, \citenamefont {Koner}, \citenamefont
  {Tomza}, \citenamefont {Meuwly},\ and\ \citenamefont
  {Willitsch}}]{doerfler19a}%
  \BibitemOpen
  \bibfield  {author} {\bibinfo {author} {\bibfnamefont {A.~D.}\ \bibnamefont
  {D{\"o}rfler}}, \bibinfo {author} {\bibfnamefont {P.}~\bibnamefont {Eberle}},
  \bibinfo {author} {\bibfnamefont {D.}~\bibnamefont {Koner}}, \bibinfo
  {author} {\bibfnamefont {M.}~\bibnamefont {Tomza}}, \bibinfo {author}
  {\bibfnamefont {M.}~\bibnamefont {Meuwly}}, \ and\ \bibinfo {author}
  {\bibfnamefont {S.}~\bibnamefont {Willitsch}},\ }\href@noop {} {\bibfield
  {journal} {\bibinfo  {journal} {arXiv preprint arXiv:1906.12285}\ } (\bibinfo
  {year} {2019})}\BibitemShut {NoStop}%
\bibitem [{\citenamefont {Harty}\ \emph {et~al.}(2014)\citenamefont {Harty},
  \citenamefont {Allcock}, \citenamefont {Ballance}, \citenamefont {Guidoni},
  \citenamefont {Janacek}, \citenamefont {Linke}, \citenamefont {Stacey},\ and\
  \citenamefont {Lucas}}]{harty14a}%
  \BibitemOpen
  \bibfield  {author} {\bibinfo {author} {\bibfnamefont {T.~P.}\ \bibnamefont
  {Harty}}, \bibinfo {author} {\bibfnamefont {D.~T.~C.}\ \bibnamefont
  {Allcock}}, \bibinfo {author} {\bibfnamefont {C.~J.}\ \bibnamefont
  {Ballance}}, \bibinfo {author} {\bibfnamefont {L.}~\bibnamefont {Guidoni}},
  \bibinfo {author} {\bibfnamefont {H.~A.}\ \bibnamefont {Janacek}}, \bibinfo
  {author} {\bibfnamefont {N.~M.}\ \bibnamefont {Linke}}, \bibinfo {author}
  {\bibfnamefont {D.~N.}\ \bibnamefont {Stacey}}, \ and\ \bibinfo {author}
  {\bibfnamefont {D.~M.}\ \bibnamefont {Lucas}},\ }\href@noop {} {\bibfield
  {journal} {\bibinfo  {journal} {Phys. Rev. Lett.}\ }\textbf {\bibinfo
  {volume} {113}},\ \bibinfo {pages} {220501} (\bibinfo {year}
  {2014})}\BibitemShut {NoStop}%
\bibitem [{\citenamefont {Schmidt}\ \emph {et~al.}(2005)\citenamefont
  {Schmidt}, \citenamefont {Rosenband}, \citenamefont {Langer}, \citenamefont
  {Itano}, \citenamefont {Bergquist},\ and\ \citenamefont
  {Wineland}}]{schmidt05a}%
  \BibitemOpen
  \bibfield  {author} {\bibinfo {author} {\bibfnamefont {P.~O.}\ \bibnamefont
  {Schmidt}}, \bibinfo {author} {\bibfnamefont {T.}~\bibnamefont {Rosenband}},
  \bibinfo {author} {\bibfnamefont {C.}~\bibnamefont {Langer}}, \bibinfo
  {author} {\bibfnamefont {W.~M.}\ \bibnamefont {Itano}}, \bibinfo {author}
  {\bibfnamefont {J.~C.}\ \bibnamefont {Bergquist}}, \ and\ \bibinfo {author}
  {\bibfnamefont {D.~J.}\ \bibnamefont {Wineland}},\ }\href@noop {} {\bibfield
  {journal} {\bibinfo  {journal} {Science}\ }\textbf {\bibinfo {volume}
  {309}},\ \bibinfo {pages} {749} (\bibinfo {year} {2005})}\BibitemShut
  {NoStop}%
\bibitem [{\citenamefont {Wolf}\ \emph {et~al.}(2016)\citenamefont {Wolf},
  \citenamefont {Wan}, \citenamefont {Heip}, \citenamefont {Gebert},
  \citenamefont {Shi},\ and\ \citenamefont {Schmidt}}]{wolf16a}%
  \BibitemOpen
  \bibfield  {author} {\bibinfo {author} {\bibfnamefont {F.}~\bibnamefont
  {Wolf}}, \bibinfo {author} {\bibfnamefont {Y.}~\bibnamefont {Wan}}, \bibinfo
  {author} {\bibfnamefont {J.~C.}\ \bibnamefont {Heip}}, \bibinfo {author}
  {\bibfnamefont {F.}~\bibnamefont {Gebert}}, \bibinfo {author} {\bibfnamefont
  {C.}~\bibnamefont {Shi}}, \ and\ \bibinfo {author} {\bibfnamefont {P.~O.}\
  \bibnamefont {Schmidt}},\ }\href@noop {} {\bibfield  {journal} {\bibinfo
  {journal} {Nature}\ }\textbf {\bibinfo {volume} {530}},\ \bibinfo {pages}
  {457} (\bibinfo {year} {2016})}\BibitemShut {NoStop}%
\bibitem [{\citenamefont {Chou}\ \emph {et~al.}(2017)\citenamefont {Chou},
  \citenamefont {Kurz}, \citenamefont {Hume}, \citenamefont {Plessow},
  \citenamefont {Leibrandt},\ and\ \citenamefont {Leibfried}}]{chou17a}%
  \BibitemOpen
  \bibfield  {author} {\bibinfo {author} {\bibfnamefont {C.~W.}\ \bibnamefont
  {Chou}}, \bibinfo {author} {\bibfnamefont {C.}~\bibnamefont {Kurz}}, \bibinfo
  {author} {\bibfnamefont {D.~B.}\ \bibnamefont {Hume}}, \bibinfo {author}
  {\bibfnamefont {P.~N.}\ \bibnamefont {Plessow}}, \bibinfo {author}
  {\bibfnamefont {D.~R.}\ \bibnamefont {Leibrandt}}, \ and\ \bibinfo {author}
  {\bibfnamefont {D.}~\bibnamefont {Leibfried}},\ }\href@noop {} {\bibfield
  {journal} {\bibinfo  {journal} {Nature}\ }\textbf {\bibinfo {volume} {545}},\
  \bibinfo {pages} {203} (\bibinfo {year} {2017})}\BibitemShut {NoStop}%
\bibitem [{\citenamefont {Braginsky}\ \emph {et~al.}(1980)\citenamefont
  {Braginsky}, \citenamefont {Vorontsov},\ and\ \citenamefont
  {Thorne}}]{braginsky80}%
  \BibitemOpen
  \bibfield  {author} {\bibinfo {author} {\bibfnamefont {V.~B.}\ \bibnamefont
  {Braginsky}}, \bibinfo {author} {\bibfnamefont {Y.~I.}\ \bibnamefont
  {Vorontsov}}, \ and\ \bibinfo {author} {\bibfnamefont {K.~S.}\ \bibnamefont
  {Thorne}},\ }\href@noop {} {\bibfield  {journal} {\bibinfo  {journal}
  {Science}\ }\textbf {\bibinfo {volume} {209}},\ \bibinfo {pages} {547}
  (\bibinfo {year} {1980})}\BibitemShut {NoStop}%
\bibitem [{\citenamefont {Braginsky}\ and\ \citenamefont
  {Khalili}(1996)}]{braginsky96}%
  \BibitemOpen
  \bibfield  {author} {\bibinfo {author} {\bibfnamefont {V.~B.}\ \bibnamefont
  {Braginsky}}\ and\ \bibinfo {author} {\bibfnamefont {F.~Y.}\ \bibnamefont
  {Khalili}},\ }\href@noop {} {\bibfield  {journal} {\bibinfo  {journal} {Rev.
  Mod. Phys.}\ }\textbf {\bibinfo {volume} {68}},\ \bibinfo {pages} {1}
  (\bibinfo {year} {1996})}\BibitemShut {NoStop}%
\bibitem [{\citenamefont {Hume}\ \emph {et~al.}(2007)\citenamefont {Hume},
  \citenamefont {Rosenband},\ and\ \citenamefont {Wineland}}]{hume07a}%
  \BibitemOpen
  \bibfield  {author} {\bibinfo {author} {\bibfnamefont {D.}~\bibnamefont
  {Hume}}, \bibinfo {author} {\bibfnamefont {T.}~\bibnamefont {Rosenband}}, \
  and\ \bibinfo {author} {\bibfnamefont {D.~J.}\ \bibnamefont {Wineland}},\
  }\href@noop {} {\bibfield  {journal} {\bibinfo  {journal} {Phys. Rev. Lett.}\
  }\textbf {\bibinfo {volume} {99}},\ \bibinfo {pages} {120502} (\bibinfo
  {year} {2007})}\BibitemShut {NoStop}%
\bibitem [{\citenamefont {Kajita}(2015)}]{kajita15a}%
  \BibitemOpen
  \bibfield  {author} {\bibinfo {author} {\bibfnamefont {M.}~\bibnamefont
  {Kajita}},\ }\href@noop {} {\bibfield  {journal} {\bibinfo  {journal} {{Phys.
  Rev. A}}\ }\textbf {\bibinfo {volume} {92}},\ \bibinfo {pages} {043423}
  (\bibinfo {year} {2015})}\BibitemShut {NoStop}%
\bibitem [{\citenamefont {Kajita}\ \emph {et~al.}(2014)\citenamefont {Kajita},
  \citenamefont {Gopakumar}, \citenamefont {Abe}, \citenamefont {Hada},\ and\
  \citenamefont {Keller}}]{kajita14a}%
  \BibitemOpen
  \bibfield  {author} {\bibinfo {author} {\bibfnamefont {M.}~\bibnamefont
  {Kajita}}, \bibinfo {author} {\bibfnamefont {G.}~\bibnamefont {Gopakumar}},
  \bibinfo {author} {\bibfnamefont {M.}~\bibnamefont {Abe}}, \bibinfo {author}
  {\bibfnamefont {M.}~\bibnamefont {Hada}}, \ and\ \bibinfo {author}
  {\bibfnamefont {M.}~\bibnamefont {Keller}},\ }\href@noop {} {\bibfield
  {journal} {\bibinfo  {journal} {{Phys. Rev. A}}\ }\textbf {\bibinfo {volume}
  {89}},\ \bibinfo {pages} {032509} (\bibinfo {year} {2014})}\BibitemShut
  {NoStop}%
\bibitem [{\citenamefont {Schiller}\ \emph {et~al.}(2014)\citenamefont
  {Schiller}, \citenamefont {Bakalov},\ and\ \citenamefont
  {Korobov}}]{schiller14a}%
  \BibitemOpen
  \bibfield  {author} {\bibinfo {author} {\bibfnamefont {S.}~\bibnamefont
  {Schiller}}, \bibinfo {author} {\bibfnamefont {D.}~\bibnamefont {Bakalov}}, \
  and\ \bibinfo {author} {\bibfnamefont {V.~I.}\ \bibnamefont {Korobov}},\
  }\href@noop {} {\bibfield  {journal} {\bibinfo  {journal} {{Phys. Rev.
  Lett.}}\ }\textbf {\bibinfo {volume} {113}},\ \bibinfo {pages} {023004}
  (\bibinfo {year} {2014})}\BibitemShut {NoStop}%
\bibitem [{\citenamefont {Kimble}(2008)}]{kimble08}%
  \BibitemOpen
  \bibfield  {author} {\bibinfo {author} {\bibfnamefont {H.~J.}\ \bibnamefont
  {Kimble}},\ }\href@noop {} {\bibfield  {journal} {\bibinfo  {journal}
  {Nature}\ }\textbf {\bibinfo {volume} {453}},\ \bibinfo {pages} {1023}
  (\bibinfo {year} {2008})}\BibitemShut {NoStop}%
\bibitem [{\citenamefont {Wehner}\ \emph {et~al.}(2018)\citenamefont {Wehner},
  \citenamefont {Elkouss},\ and\ \citenamefont {Hanson}}]{wehner18}%
  \BibitemOpen
  \bibfield  {author} {\bibinfo {author} {\bibfnamefont {S.}~\bibnamefont
  {Wehner}}, \bibinfo {author} {\bibfnamefont {D.}~\bibnamefont {Elkouss}}, \
  and\ \bibinfo {author} {\bibfnamefont {R.}~\bibnamefont {Hanson}},\
  }\href@noop {} {\bibfield  {journal} {\bibinfo  {journal} {Science}\ }\textbf
  {\bibinfo {volume} {362}},\ \bibinfo {pages} {eaam9288} (\bibinfo {year}
  {2018})}\BibitemShut {NoStop}%
\bibitem [{\citenamefont {Meir}\ \emph {et~al.}(2019)\citenamefont {Meir},
  \citenamefont {Hegi}, \citenamefont {Najafian}, \citenamefont {Sinhal},\ and\
  \citenamefont {Willitsch}}]{meir19a}%
  \BibitemOpen
  \bibfield  {author} {\bibinfo {author} {\bibfnamefont {Z.}~\bibnamefont
  {Meir}}, \bibinfo {author} {\bibfnamefont {G.}~\bibnamefont {Hegi}}, \bibinfo
  {author} {\bibfnamefont {K.}~\bibnamefont {Najafian}}, \bibinfo {author}
  {\bibfnamefont {M.}~\bibnamefont {Sinhal}}, \ and\ \bibinfo {author}
  {\bibfnamefont {S.}~\bibnamefont {Willitsch}},\ }\href@noop {} {\bibfield
  {journal} {\bibinfo  {journal} {Faraday Discuss.}\ }\textbf {\bibinfo
  {volume} {217}},\ \bibinfo {pages} {561} (\bibinfo {year}
  {2019})}\BibitemShut {NoStop}%
\bibitem [{\citenamefont {Hume}\ \emph {et~al.}(2011)\citenamefont {Hume},
  \citenamefont {Chou}, \citenamefont {Leibrandt}, \citenamefont {Thorpe},
  \citenamefont {Wineland},\ and\ \citenamefont {Rosenband}}]{hume11a}%
  \BibitemOpen
  \bibfield  {author} {\bibinfo {author} {\bibfnamefont {D.}~\bibnamefont
  {Hume}}, \bibinfo {author} {\bibfnamefont {C.~W.}\ \bibnamefont {Chou}},
  \bibinfo {author} {\bibfnamefont {D.~R.}\ \bibnamefont {Leibrandt}}, \bibinfo
  {author} {\bibfnamefont {M.~J.}\ \bibnamefont {Thorpe}}, \bibinfo {author}
  {\bibfnamefont {D.~J.}\ \bibnamefont {Wineland}}, \ and\ \bibinfo {author}
  {\bibfnamefont {T.}~\bibnamefont {Rosenband}},\ }\href@noop {} {\bibfield
  {journal} {\bibinfo  {journal} {Phys. Rev. Lett.}\ }\textbf {\bibinfo
  {volume} {107}},\ \bibinfo {pages} {243902} (\bibinfo {year}
  {2011})}\BibitemShut {NoStop}%
\bibitem [{\citenamefont {Koelemeij}\ \emph {et~al.}(2007)\citenamefont
  {Koelemeij}, \citenamefont {Roth},\ and\ \citenamefont
  {Schiller}}]{koelemeij07a}%
  \BibitemOpen
  \bibfield  {author} {\bibinfo {author} {\bibfnamefont {J.~C.}\ \bibnamefont
  {Koelemeij}}, \bibinfo {author} {\bibfnamefont {B.}~\bibnamefont {Roth}}, \
  and\ \bibinfo {author} {\bibfnamefont {S.}~\bibnamefont {Schiller}},\
  }\href@noop {} {\bibfield  {journal} {\bibinfo  {journal} {{Phys. Rev. A}}\
  }\textbf {\bibinfo {volume} {76}},\ \bibinfo {pages} {023413} (\bibinfo
  {year} {2007})}\BibitemShut {NoStop}%
\bibitem [{\citenamefont {Christensen}\ \emph {et~al.}(2019)\citenamefont
  {Christensen}, \citenamefont {Hucul}, \citenamefont {Campbell},\ and\
  \citenamefont {Hudson}}]{christensen19}%
  \BibitemOpen
  \bibfield  {author} {\bibinfo {author} {\bibfnamefont {J.~E.}\ \bibnamefont
  {Christensen}}, \bibinfo {author} {\bibfnamefont {D.}~\bibnamefont {Hucul}},
  \bibinfo {author} {\bibfnamefont {W.~C.}\ \bibnamefont {Campbell}}, \ and\
  \bibinfo {author} {\bibfnamefont {E.~R.}\ \bibnamefont {Hudson}},\
  }\href@noop {} {\bibfield  {journal} {\bibinfo  {journal} {arXiv preprint
  arXiv:1907.13331}\ } (\bibinfo {year} {2019})}\BibitemShut {NoStop}%
\bibitem [{\citenamefont {Biercuk}\ \emph {et~al.}(2010)\citenamefont
  {Biercuk}, \citenamefont {Uys}, \citenamefont {Britton}, \citenamefont
  {VanDevender},\ and\ \citenamefont {Bollinger}}]{biercuk10}%
  \BibitemOpen
  \bibfield  {author} {\bibinfo {author} {\bibfnamefont {M.~J.}\ \bibnamefont
  {Biercuk}}, \bibinfo {author} {\bibfnamefont {H.}~\bibnamefont {Uys}},
  \bibinfo {author} {\bibfnamefont {J.~W.}\ \bibnamefont {Britton}}, \bibinfo
  {author} {\bibfnamefont {A.~P.}\ \bibnamefont {VanDevender}}, \ and\ \bibinfo
  {author} {\bibfnamefont {J.~J.}\ \bibnamefont {Bollinger}},\ }\href@noop {}
  {\bibfield  {journal} {\bibinfo  {journal} {Nat. Nanotechnol.}\ }\textbf
  {\bibinfo {volume} {5}},\ \bibinfo {pages} {646} (\bibinfo {year}
  {2010})}\BibitemShut {NoStop}%
\bibitem [{\citenamefont {Gardner}\ \emph {et~al.}(2019)\citenamefont
  {Gardner}, \citenamefont {Softley},\ and\ \citenamefont
  {Keller}}]{gardner19a}%
  \BibitemOpen
  \bibfield  {author} {\bibinfo {author} {\bibfnamefont {A.}~\bibnamefont
  {Gardner}}, \bibinfo {author} {\bibfnamefont {T.}~\bibnamefont {Softley}}, \
  and\ \bibinfo {author} {\bibfnamefont {M.}~\bibnamefont {Keller}},\
  }\href@noop {} {\bibfield  {journal} {\bibinfo  {journal} {Sci. Rep.}\
  }\textbf {\bibinfo {volume} {9}},\ \bibinfo {pages} {506} (\bibinfo {year}
  {2019})}\BibitemShut {NoStop}%
\bibitem [{\citenamefont {Meekhof}\ \emph {et~al.}(1996)\citenamefont
  {Meekhof}, \citenamefont {Monroe}, \citenamefont {King}, \citenamefont
  {Itano},\ and\ \citenamefont {Wineland}}]{meekhof96a}%
  \BibitemOpen
  \bibfield  {author} {\bibinfo {author} {\bibfnamefont {D.~M.}\ \bibnamefont
  {Meekhof}}, \bibinfo {author} {\bibfnamefont {C.}~\bibnamefont {Monroe}},
  \bibinfo {author} {\bibfnamefont {B.~E.}\ \bibnamefont {King}}, \bibinfo
  {author} {\bibfnamefont {W.~M.}\ \bibnamefont {Itano}}, \ and\ \bibinfo
  {author} {\bibfnamefont {D.~J.}\ \bibnamefont {Wineland}},\ }\href@noop {}
  {\bibfield  {journal} {\bibinfo  {journal} {Phys. Rev. Lett.}\ }\textbf
  {\bibinfo {volume} {76}},\ \bibinfo {pages} {1796} (\bibinfo {year}
  {1996})}\BibitemShut {NoStop}%
\bibitem [{\citenamefont {Leibfried}\ \emph {et~al.}(2003)\citenamefont
  {Leibfried}, \citenamefont {Blatt}, \citenamefont {Monroe},\ and\
  \citenamefont {Wineland}}]{leibfried03a}%
  \BibitemOpen
  \bibfield  {author} {\bibinfo {author} {\bibfnamefont {D.}~\bibnamefont
  {Leibfried}}, \bibinfo {author} {\bibfnamefont {R.}~\bibnamefont {Blatt}},
  \bibinfo {author} {\bibfnamefont {C.}~\bibnamefont {Monroe}}, \ and\ \bibinfo
  {author} {\bibfnamefont {D.}~\bibnamefont {Wineland}},\ }\href@noop {}
  {\bibfield  {journal} {\bibinfo  {journal} {Rev. Mod. Phys.}\ }\textbf
  {\bibinfo {volume} {75}},\ \bibinfo {pages} {281} (\bibinfo {year}
  {2003})}\BibitemShut {NoStop}%
\bibitem [{\citenamefont {Herzberg}(1991)}]{herzberg91a}%
  \BibitemOpen
  \bibfield  {author} {\bibinfo {author} {\bibfnamefont {G.}~\bibnamefont
  {Herzberg}},\ }\href@noop {} {\emph {\bibinfo {title} {{Molecular Spectra and
  Molecular Structure, Volume II, Infrared and Raman Spectra of Polyatomic
  Molecules}}}}\ (\bibinfo  {publisher} {Krieger},\ \bibinfo {address}
  {Malabar},\ \bibinfo {year} {1991})\BibitemShut {NoStop}%
\bibitem [{\citenamefont {Wu}\ \emph {et~al.}(2007)\citenamefont {Wu},
  \citenamefont {Ben}, \citenamefont {Li}, \citenamefont {Zheng}, \citenamefont
  {Chen},\ and\ \citenamefont {Yang}}]{wu07a}%
  \BibitemOpen
  \bibfield  {author} {\bibinfo {author} {\bibfnamefont {Y.-D.}\ \bibnamefont
  {Wu}}, \bibinfo {author} {\bibfnamefont {J.-W.}\ \bibnamefont {Ben}},
  \bibinfo {author} {\bibfnamefont {B.}~\bibnamefont {Li}}, \bibinfo {author}
  {\bibfnamefont {L.-J.}\ \bibnamefont {Zheng}}, \bibinfo {author}
  {\bibfnamefont {Y.-Q.}\ \bibnamefont {Chen}}, \ and\ \bibinfo {author}
  {\bibfnamefont {X.-H.}\ \bibnamefont {Yang}},\ }\href@noop {} {\bibfield
  {journal} {\bibinfo  {journal} {Chin. J. Chem. Phys.}\ }\textbf {\bibinfo
  {volume} {20}},\ \bibinfo {pages} {285} (\bibinfo {year} {2007})}\BibitemShut
  {NoStop}%
\bibitem [{\citenamefont {Bachir}\ \emph {et~al.}(1994)\citenamefont {Bachir},
  \citenamefont {Bolvin}, \citenamefont {Demuynck}, \citenamefont {Destombes},\
  and\ \citenamefont {Zellagui}}]{bachir94a}%
  \BibitemOpen
  \bibfield  {author} {\bibinfo {author} {\bibfnamefont {I.~H.}\ \bibnamefont
  {Bachir}}, \bibinfo {author} {\bibfnamefont {H.}~\bibnamefont {Bolvin}},
  \bibinfo {author} {\bibfnamefont {C.}~\bibnamefont {Demuynck}}, \bibinfo
  {author} {\bibfnamefont {J.}~\bibnamefont {Destombes}}, \ and\ \bibinfo
  {author} {\bibfnamefont {A.}~\bibnamefont {Zellagui}},\ }\href@noop {}
  {\bibfield  {journal} {\bibinfo  {journal} {J. Mol. Spectrosc.}\ }\textbf
  {\bibinfo {volume} {166}},\ \bibinfo {pages} {88} (\bibinfo {year}
  {1994})}\BibitemShut {NoStop}%
\bibitem [{\citenamefont {Harada}\ \emph {et~al.}(1994)\citenamefont {Harada},
  \citenamefont {Wada},\ and\ \citenamefont {Tanaka}}]{harada94a}%
  \BibitemOpen
  \bibfield  {author} {\bibinfo {author} {\bibfnamefont {K.}~\bibnamefont
  {Harada}}, \bibinfo {author} {\bibfnamefont {T.}~\bibnamefont {Wada}}, \ and\
  \bibinfo {author} {\bibfnamefont {T.}~\bibnamefont {Tanaka}},\ }\href@noop {}
  {\bibfield  {journal} {\bibinfo  {journal} {J. Mol. Spectrosc.}\ }\textbf
  {\bibinfo {volume} {163}},\ \bibinfo {pages} {436} (\bibinfo {year}
  {1994})}\BibitemShut {NoStop}%
\bibitem [{\citenamefont {Cartwright}(1973)}]{cartwright73a}%
  \BibitemOpen
  \bibfield  {author} {\bibinfo {author} {\bibfnamefont {D.~C.}\ \bibnamefont
  {Cartwright}},\ }\href@noop {} {\bibfield  {journal} {\bibinfo  {journal} {J.
  Chem. Phys.}\ }\textbf {\bibinfo {volume} {58}},\ \bibinfo {pages} {178}
  (\bibinfo {year} {1973})}\BibitemShut {NoStop}%
\bibitem [{\citenamefont {Langhoff}\ \emph {et~al.}(1987)\citenamefont
  {Langhoff}, \citenamefont {Bauschlicher~Jr},\ and\ \citenamefont
  {Partridge}}]{langhoff87}%
  \BibitemOpen
  \bibfield  {author} {\bibinfo {author} {\bibfnamefont {S.~R.}\ \bibnamefont
  {Langhoff}}, \bibinfo {author} {\bibfnamefont {C.~W.}\ \bibnamefont
  {Bauschlicher~Jr}}, \ and\ \bibinfo {author} {\bibfnamefont {H.}~\bibnamefont
  {Partridge}},\ }\href@noop {} {\bibfield  {journal} {\bibinfo  {journal} {J.
  Chem. Phys.}\ }\textbf {\bibinfo {volume} {87}},\ \bibinfo {pages} {4716}
  (\bibinfo {year} {1987})}\BibitemShut {NoStop}%
\bibitem [{\citenamefont {Gilmore}\ \emph {et~al.}(1992)\citenamefont
  {Gilmore}, \citenamefont {Laher},\ and\ \citenamefont {Espy}}]{gilmore92}%
  \BibitemOpen
  \bibfield  {author} {\bibinfo {author} {\bibfnamefont {F.~R.}\ \bibnamefont
  {Gilmore}}, \bibinfo {author} {\bibfnamefont {R.~R.}\ \bibnamefont {Laher}},
  \ and\ \bibinfo {author} {\bibfnamefont {P.~J.}\ \bibnamefont {Espy}},\
  }\href@noop {} {\bibfield  {journal} {\bibinfo  {journal} {J. Phys. Chem.
  Ref. Data}\ }\textbf {\bibinfo {volume} {21}},\ \bibinfo {pages} {1005}
  (\bibinfo {year} {1992})}\BibitemShut {NoStop}%
\bibitem [{\citenamefont {Bruna}\ and\ \citenamefont {Grein}(2008)}]{bruna08}%
  \BibitemOpen
  \bibfield  {author} {\bibinfo {author} {\bibfnamefont {P.~J.}\ \bibnamefont
  {Bruna}}\ and\ \bibinfo {author} {\bibfnamefont {F.}~\bibnamefont {Grein}},\
  }\href@noop {} {\bibfield  {journal} {\bibinfo  {journal} {J. Mol.
  Spectrosc.}\ }\textbf {\bibinfo {volume} {250}},\ \bibinfo {pages} {75}
  (\bibinfo {year} {2008})}\BibitemShut {NoStop}%
\bibitem [{\citenamefont {Zhou}\ \emph {et~al.}(2010)\citenamefont {Zhou},
  \citenamefont {Xu}, \citenamefont {Chen},\ and\ \citenamefont
  {Chen}}]{zhou10}%
  \BibitemOpen
  \bibfield  {author} {\bibinfo {author} {\bibfnamefont {X.}~\bibnamefont
  {Zhou}}, \bibinfo {author} {\bibfnamefont {X.}~\bibnamefont {Xu}}, \bibinfo
  {author} {\bibfnamefont {X.}~\bibnamefont {Chen}}, \ and\ \bibinfo {author}
  {\bibfnamefont {J.}~\bibnamefont {Chen}},\ }\href@noop {} {\bibfield
  {journal} {\bibinfo  {journal} {Phys. Rev. A}\ }\textbf {\bibinfo {volume}
  {81}},\ \bibinfo {pages} {012115} (\bibinfo {year} {2010})}\BibitemShut
  {NoStop}%
\bibitem [{\citenamefont {Bennett}(1970)}]{bennett70}%
  \BibitemOpen
  \bibfield  {author} {\bibinfo {author} {\bibfnamefont {R.}~\bibnamefont
  {Bennett}},\ }\href@noop {} {\bibfield  {journal} {\bibinfo  {journal} {Mon.
  Not. R. Astron. Soc.}\ }\textbf {\bibinfo {volume} {147}},\ \bibinfo {pages}
  {35} (\bibinfo {year} {1970})}\BibitemShut {NoStop}%
\bibitem [{\citenamefont {Grimm}\ \emph {et~al.}(2000)\citenamefont {Grimm},
  \citenamefont {Weidem{\"u}ller},\ and\ \citenamefont
  {Ovchinnikov}}]{grimm00a}%
  \BibitemOpen
  \bibfield  {author} {\bibinfo {author} {\bibfnamefont {R.}~\bibnamefont
  {Grimm}}, \bibinfo {author} {\bibfnamefont {M.}~\bibnamefont
  {Weidem{\"u}ller}}, \ and\ \bibinfo {author} {\bibfnamefont {Y.~B.}\
  \bibnamefont {Ovchinnikov}},\ }\href@noop {} {\bibfield  {journal} {\bibinfo
  {journal} {Adv. At. Mol. Opt. Phys.}\ }\textbf {\bibinfo {volume} {42}},\
  \bibinfo {pages} {95} (\bibinfo {year} {2000})}\BibitemShut {NoStop}%
\bibitem [{\citenamefont {Germann}(2016)}]{germann16d}%
  \BibitemOpen
  \bibfield  {author} {\bibinfo {author} {\bibfnamefont {M.}~\bibnamefont
  {Germann}},\ }\href@noop {} {Ph.D. thesis},\ \bibinfo  {school} {University
  of Basel} (\bibinfo {year} {2016})\BibitemShut {NoStop}%
\bibitem [{\citenamefont {Morigi}\ and\ \citenamefont
  {Walther}(2001)}]{morigi01a}%
  \BibitemOpen
  \bibfield  {author} {\bibinfo {author} {\bibfnamefont {G.}~\bibnamefont
  {Morigi}}\ and\ \bibinfo {author} {\bibfnamefont {H.}~\bibnamefont
  {Walther}},\ }\href@noop {} {\bibfield  {journal} {\bibinfo  {journal} {Eur.
  Phys. J. D}\ }\textbf {\bibinfo {volume} {13}},\ \bibinfo {pages} {261}
  (\bibinfo {year} {2001})}\BibitemShut {NoStop}%
\end{thebibliography}
\bibliographystyle{apsrev4-1}

\newpage{}
\onecolumngrid

\section*{Supplementary materials}

\subsection*{Ac-Stark shift calculation}
We calculate the ac-Stark shift of the \NN{} molecule within second-order perturbation theory outside the rotating-wave approximation for linearly $\pi$-polarized light \cite{zhou10}, 
\begin{equation}
    \Delta E_i = -\sum_j \frac{3 \pi c^2}{\omega_{ij}^2 \left(\omega_{ij}^2-\omega^2\right)}\cdot I \cdot S_\textrm{rot}^{ij} \cdot A_\textrm{vib}^{ij} \left|\sqrt{2J_j+1}
    \begin{pmatrix}
        J_j & 1 & J_i \\
        -m & 0 & m 
    \end{pmatrix}
    \right|^2.
\end{equation}\label{eq:acshift}
Here, the ac-Stark shift $\Delta E_i$ of the state \ket{i} is given by the sum over all contributions from dipole allowed transitions to excited levels \ket{j}, where $\omega_{ij}$ are the angular transition frequencies \cite{wu07a}, $\omega$ is the angular laser frequency, $I$ is the laser intensity, $A_\textrm{vib}^{ij}$ is the vibronic Einstein $A$-coefficient which includes the Franck-Condon factor \cite{langhoff87}, $S_\textrm{rot}^{ij}$ is the normalized H\"onl-London factor \cite{bennett70}, $J$ is the total angular momentum without nuclear spin and $m$ is the projection of the angular momentum on the quantization axis defined by the external magnetic field. The value of $m$ is shared between the states due to the $\pi$ polarization of the laser used here. 

For the state \ket{i} = \NNdown, in the frequency region of Fig. \ref{fig:N2_ODF} of the main text, the ac-Stark shift is dominated by the $R_{11}(1/2)$ transition to the $A^2\Pi_{u}(v'=2)$ electronic state of \NN. Moreover, since for this state there is no rotational or electronic angular momentum, the ac-Stark shift is not affected significantly by hyperfine interactions \cite{grimm00a} for $\Delta\gg\Delta_\textrm{hfs}$ where $\Delta_\textrm{hfs}\approx$ 300 MHz is the assumed spacing of the hyperfine manifold of the transition. The hyperfine structure of the $A^2\Pi_{u}-X^2\Sigma_{g}^+$ band in \NN{} was not experimentally measured, to the best of our knowledge, and we give here typical values for transitions within the $X^2\Sigma_{g}^+$ state \cite{germann16d} as an order-of-magnitude estimate. For states with rotational angular momentum $N>0$, the ac-Stark shift depends on the specific fine, hyperfine and Zeeman state.

\subsection*{D-state purification}
We initialize the \Ca{} ion in the \Caup{} = \ket{D_{5/2}(m=-5/2)} state since it almost decouples completely from the lattice beams due to their $\pi$ polarization whereas the \Cadown=\ket{S_{1/2}(m=-1/2)} couples through the excited $P_{1/2}$ and $P_{3/2}$ states \cite{meir19a}. To do so, we use a $\pi$-pulse which typically transfers 97\% of the population to the \Caup{} state. The remaining 3\% population in the \Cadown{} state, due to imperfect ground-state cooling and pulse parameters, introduces background noise in the state-detection measurement. To reduce this background, we apply a detection pulse on the $S_{1/2}\leftrightarrow P_{1/2}\leftrightarrow D_{3/2}$ closed cycling transition prior to the ODF pulse and post-select only the experiments in which the detection resulted in an ion in the $D$ state \cite{chou17a}. Experiments in which the ion was projected to the S state were excluded from the analysis. 

\subsection*{Fidelity of the state detection}
The probability to get $k$ successes in $N$ QND measurements, i.e. $k$ projections to the \Cadown{} state after $N$ BSB pulses, follows a binomial distribution, $B(N,p)$, where $p$=$P$(\Cadown{}) is the probability of a successful excitation with a BSB pulse. Thus, the likelihood of a BSB success probability, $p$, given $k$ successes in $N$ QND measurements is given by,
\begin{equation}\label{eq:defL}
    L(p|k,N)=\frac{n!}{k!(N-k)!}p^k(1-p)^{n-k}.
\end{equation}
Given two molecular states with two different BSB success probabilities, $p_\alpha>p_\beta$, the threshold, $k_t$, for discriminating between them can be calculated by solving,
\begin{equation}
    k_t=\max_k \left[ L(p_\alpha|k,N)<L(p_\beta|k,N) \right].
\end{equation}
Here, for $k\leq k_t$ the state is determined as ``dark'' while for $k>k_t$ the state is determined as ``bright''. This equation can be solved analytically by using Eq. \ref{eq:defL} and taking the logarithm of both sides yielding,
\begin{equation}
    k_t=\left \lfloor {\frac{N}{\frac{\log [p_\alpha/p_\beta]}{\log [(1-p_\beta)/(1-p_\alpha)]} + 1}
    }\right \rfloor.
\end{equation}
For the parameters used in the experiment ($p_\alpha$=0.52, $p_\beta$=0.06 and $N$=22), $k_t$=5 such that for $P$(\Cadown{})$<$0.23 the state is ``dark'' while for $P$(\Cadown{})$>$0.27 the state is ``bright'' hence the threshold was determined, for clarity, to be 0.25.

We can use this (or any other) threshold to calculate the ``dark'' and ``bright'' errors, $\mathcal{E}_d$ and $\mathcal{E}_b$, respectively, for detecting a ``dark'' molecule as ``bright'' and vice-versa,
\begin{equation}
    \mathcal{E}_d=\sum_{k=k_t+1}^N L(p_{\beta}|k,N),
\end{equation}
\begin{equation}
    \mathcal{E}_b=\sum_{k=0}^{k_t} L(p_{\alpha}|k,N).
\end{equation}
For the parameters used in the experiment, $\mathcal{E}_d=1.5\cdot10^{-3}$ and $\mathcal{E}_b=4.7\cdot10^{-3}$ such the theoretical fidelity of our QND measurement is calculated to be $\mathcal{F}\approx99.5$\%. 

\subsection*{Extracting molecular ac-Stark shifts from blue-sideband Rabi flops}

\begin{figure}
	\centering
	\includegraphics[width=0.8\linewidth]{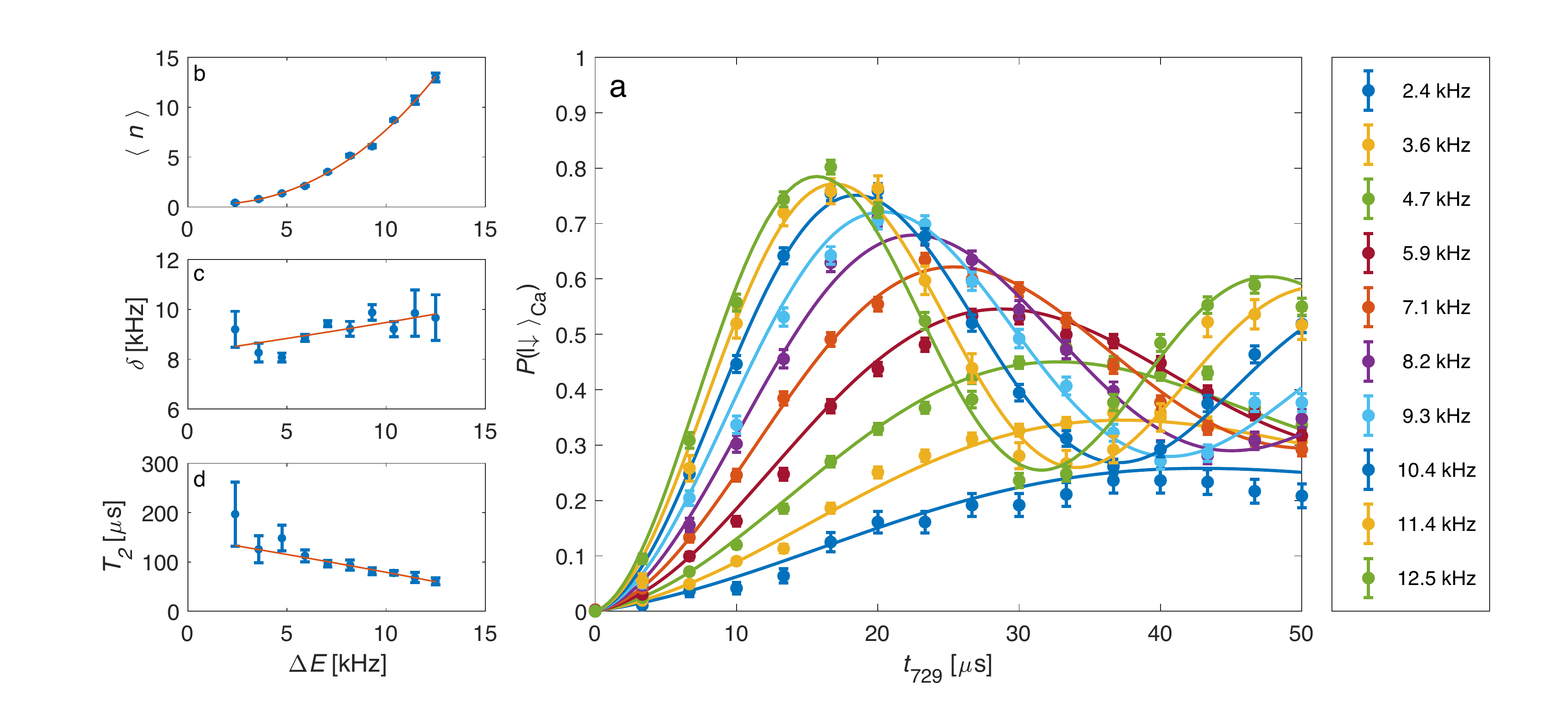}
 	\caption{\textbf{Ac-Stark shift calibration.} a) Red-sideband oscillations due to a pre-determined modulated ac-Stark shift amplitude applied on \Ca{} when \NN{} experiences no ac-Stark shift. Lines are fits to the function defined in Eq. \ref{eq:fitfun1} and \ref{eq:fitfun2}. The ac-Stark shift experienced by the molecule is given in the legend. b-d) Polynomial fits (red lines) of the parameters (blue points) extracted from fits to the RSB oscillations in (a). Error bars represent 68\% confidence intervals.}
	\label{fig:Ca_simulation}
\end{figure}

In Fig. \ref{fig:ForceSpectra} of the main text, we quoted the amplitude of the ac-Stark shift experienced by \NN{} derived from a BSB Rabi oscillation signal. In principle, the BSB signal can be calculated analytically knowing the Fock-state distribution \cite{leibfried03a} which is expected for the chosen ODF parameters \cite{meir19a}. However, due to experimental imperfections and decoherence, the underlying Fock-state distribution may deviate from theoretical predictions. Thus, we took a direct approach and experimentally simulated the expected BSB Rabi signal using our \Ca{}-\NN{} two-ion crystal.

For the simulations of the expected BSB Rabi signal, we performed an experiment where the \Ca{} ion experienced a modulated ac-Stark shift while the \NN{} molecular state was chosen such that its interaction with the lattice beams is negligible and thus experiences no ac-Stark shift. We initialized the \Ca{} ion in the \Cadown{} = \ket{S_{1/2}(m=-1/2)} state which couples to the lattice beams due to its interaction with the $P_{1/2}$ and $P_{3/2}$ states. We measured the magnitude of the ac-Stark shift on the narrow electric-quadrupole \Cadown{} $\leftarrow$ \Caup{} transition using a single \Ca{} ion as a probe and tuned the magnitude of the ac-Stark shift by changing the lasers power. We used the same lattice-beam duration of 500~$\mu$s to excite coherent motion in the ion crystal. Since in this experiment, opposed to the experiments described in the main text,  we prepared the system in the \Cadown{} state, we applied a red-sideband (RSB) pulse (instead of a BSB pulse) to measure Rabi oscillations. The results for various ac-shifts experienced by the \Ca{} are shown in Fig. \ref{fig:Ca_simulation}a.

We fitted the RSB flops to the general solution of RSB Rabi oscillations \cite{leibfried03a},
\begin{equation}\label{eq:fitfun1}
    y = \sum_n P(n|\alpha) \frac{\Omega_{n,n-1}^2}{\Omega_{n,n-1}^2+\delta^2}\sin^2{\left(\sqrt{\Omega_{n,n-1}^2+\delta^2} \cdot t_{729} /2 \right)},
\end{equation}
with the addition of phase decoherence characterized by a time constant $T_2$,
\begin{equation}\label{eq:fitfun2}
    P({\left| \downarrow \right\rangle_\textrm{Ca}})=ye^{-t_{729}/T_2}+\left(1-e^{-t_{729}/T_2}\right)/2.
\end{equation}
Here, $\Omega_{n,n-1}=\Omega_0 \eta e^{-\eta^2/2} L^1_{n-1}(\eta^2)/\sqrt{n}$ is the generalized Rabi frequency ($\Omega_0/2\pi\approx90$ kHz is the bare Rabi frequency, $\eta\approx0.1$ is the Lamb-Dicke parameter and $L^1_n(x)$ is a generalized Laguerre polynomial), $\delta$ is the detuning of the laser frequency from resonance with the RSB, $t_{729}$ is the RSB pulse time and $P(n|\alpha)$ is the Fock-state distribution for a given coherent state, \ket{\alpha}, which was defined in the main text.

We used three fit parameters, $\langle n \rangle=\alpha^2$, $\delta$ and $T_2$, to achieve a good agreement with the experimental data (Fig. \ref{fig:Ca_simulation}a). While the fitting parameters have physical meanings, here, we treated them as phenomenological quantities which are able to model the experimental Rabi flops with a minimum number of parameters. Nevertheless, since the measurements were performed in the same apparatus and conditions as the experimental data was taken, the measured RSB signals represent faithfully the expected molecular signal. 

We constructed a fitting function $P($\Cadown{}$|\Delta E,t_{729}$) with a single fit parameter, the ac-Stark shift, to fit the BSB oscillation data observed when the force acted on \NN{}. We did so by fitting the fit parameters of the RSB signal, $\langle n \rangle(\Delta E)$, $\delta(\Delta E)$ and $T_2(\Delta E)$, to second and first order polynomial functions as shown in Figs. \ref{fig:Ca_simulation}b-d.   

The values of the ac-Stark shifts extracted from the fitting function, $P($\Cadown{}$|\Delta E,t_{729}$), corresponds to the ac-Stark shift applied on the \Ca{} ion. However, analytic derivations and numerical simulations \cite{meir19a} of the classical Lagrangian of the system \cite{morigi01a} show that the energy of the system after applying the same force on the \NN{} is scaled by a factor due to the mass difference of the ions. To avoid this factor, the \Ca{} calibration experiments should have been performed with a laser wavelength scaled by the square-root of the ratio of the masses of the atomic ion and the molecular ion. In our case this corresponds to a wavelength of 940 nm instead of 787 nm. Instead of using a different laser for the calibration experiment, we used a classical simulation (see the appendix of Ref. \cite{meir19a} for details) to extract the correction factor for the ac-Stark shift. This correction factor increases the ac-Stark shift and the resulting vibronic-Einstein-$A$ coefficient by $\sim$17\%. 

\clearpage

\end{document}